\documentclass[preprint,showpacs,preprintnumbers,amsmath,amssymb]{revtex4}
\usepackage{xcolor}
\usepackage{graphicx}
\usepackage{dcolumn}
\usepackage{bm}
\usepackage{multirow}
\usepackage{booktabs}
\usepackage{CJK}
\usepackage{longtable}

\begin{document}

\title{Ion loss events in a cold Rb-Ca$^+$ hybrid trap: photodissociation, black-body radiation and non-radiative charge exchange}

\author{Xiaodong Xing $^{1}$, Humberto da Silva Jr $^{1,2}$, Romain Vexiau $^{1}$, Nadia Bouloufa-Maafa $^{1}$, Stefan Willitsch $^{3}$ and Olivier Dulieu $^{1,}$*}

\address{%
$^{1}$ \quad  Universit\'e Paris-Saclay, CNRS, Laboratoire Aim\'e Cotton, 91405 Orsay, France\\
$^{2}$ \quad Department of Chemistry and Biochemistry, University of Nevada, Las Vegas, 89154 Nevada, USA\\
$^{3}$ \quad Department of Chemistry, University of Basel, Klingelbergstrasse 80, 4056 Basel,  Switzerland}

\begin{abstract}
We theoretically investigate the collisional dynamics of laser-cooled $^{87}$Rb ground-state atoms and $^{40}$Ca$^+$ ground-state ions in the context of the hybrid trap experiment of Ref. [Phys. Rev. Lett. 107, 243202 (2011)], leading to ion losses. Cold $^{87}$Rb$^{40}$Ca$^+$ ground-state molecular ions are created by radiative association, and we demonstrate that they are protected against photodissociation by black-body radiation and by the $^{40}$Ca$^+$ cooling laser at 397~nm. This study yields an interpretation of the direct observation of $^{87}$Rb$^{40}$Ca$^+$ ions in the experiment, in contrast to other hybrid trap experiments using other species. Based on novel molecular data for the spin-orbit interaction, we also confirm that the non-radiative charge-exchange is the dominant loss process for Ca$^+$ and obtain rates in agreement with experimental observations and a previous calculation.
\end{abstract}

\maketitle

\section{Introduction}

In the research field of ultracold dilute matter (namely when the kinetic energy of the particles is well below $k_B \times 1$~mK and density does not exceed $10^{15}$~cm$^{-3}$, $k_B$ being the Boltzmann constant), the development of so-called hybrid traps, which merge ultracold atoms and atomic ions opened fascinating topics \cite{tomza2019}: the premises of cold chemistry with the formation of cold molecular ions \cite{hall2011, hall2012,hall2013a,hall2013b,doerfler2019,mohammadi2021}, the dynamics of ultracold inelastic collisions and charge exchange between ground-state atoms and excited-state ions involving large internal energy \cite{haze2015,saito2017,joger2017,mills2019,mingli2019,kwolek2019,benshlomi2020}, the dominant character of ultracold three-body collisions between an ion and two atoms \cite{krukow2016a,krukow2016b} due to the long-range ion-atom interaction, the enhancement of ion cooling induced by resonant charge exchange \cite{dutta2018,mahdian2021}, the emergence of ultracold ion-atom collisions in the quantum regime \cite{feldker2020} with the observation of spin-exchange \cite{sikorsky2018b} and Feshbach resonances \cite{weckesser2021}. The advantage of such atom-ion experiment is that the detection of charged particles yields reliable fingerprints to distinguish among various possible reaction channels. In all the corresponding experimental situations, the unavoidable presence of lasers to cool and trap the species has been probed to have large consequences on the hybrid trap dynamics. For instance, the detection of the above-mentioned inelastic collisions originates from the creation of excited ions during the laser-cooling process \cite{hall2011,sullivan2012}. Cold molecular ions are also expected to be mostly destroyed by these lasers \cite{jyothi2016}. 

Up to now, the direct observation of cold molecular ion has been reported in an experiment involving a magneto-optical trap (MOT) of $^{87}$Rb atoms and a Paul trap of $^{40}$Ca$^+$ \cite{hall2011,hall2013a} or $^{138}$Ba$^+$ ions \cite{hall2013b}. In these experiments, molecular ions were detected via mass spectrometry performed in the Paul trap, including RbCa$^+$, RbBa$^+$, and also Rb$_2^+$. In Ref. \cite{mohammadi2021}, the ''life and death'' of a weakly-bound RbBa$^+$ ion produced by three-body collision is inferred from the observation of the survival of a single Ba$^+$ ion embedded in an ultracold Rb quantum gas. Finally, while not observed, the formation of YbCa$^+$ \cite{rellegert2011} and BaCa$^+$ \cite{sullivan2012} ions is invoked to interpret the corresponding experimental results. 

\begin{figure}[!t]
\centering
\includegraphics[width=0.5\textwidth]{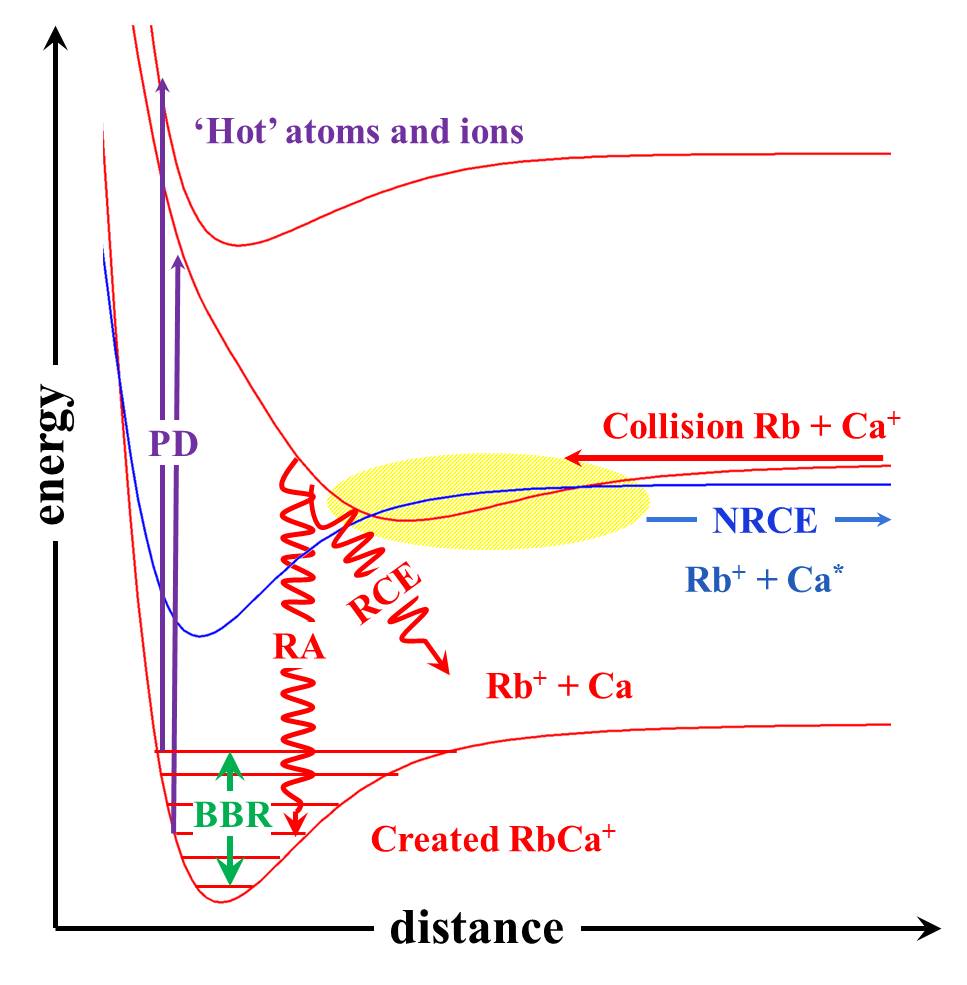}
\caption{The expected dynamical processes when a Rb atom and a Ca$^+$ ion, both in their ground state, collide at low energies (horizontal red arrow): NRCE (non radiative charge exchange) yielding a Rb$^+$ ion and an excited-state Ca$^*$ atom; RCE (radiative charge exchange) yielding a Rb$^+$ ion and a ground-state Ca atom; RA (radiative association) creating a cold ground-state molecular ion; BBR (black-body radiation) redistributing the population of the rovibrational levels of RbCa$^+$; PD (photodissociation) induced by the cooling lasers, producing excited-state Rb$^*$/Ca$^*$ atoms and  Rb$^+/$Ca$^+$ ions with outgoing energies depending on the reached dissociation threshold.}
\label{fig:processes}
\end{figure}

In order to further clarify the direct observation of $^{87}$Rb$^{40}$Ca$^+$ species in Refs. \cite{hall2011,hall2013a}, we present in this paper an extension of our previous theoretical study devoted to radiative association (RA) of laser-cooled $^{87}$Rb atoms and alkaline-earth ions \cite{dasilva2015} (hereafter referred to as paper I), invoking the processes schematized in Fig. \ref{fig:processes}. We model the photodissociation (PD) of the presumably formed ground-state $^{87}$Rb$^{40}$Ca$^+$ ions by the lasers present in the experiment of Ref. \cite{hall2011,hall2013a}. We also evaluate the role of black-body radiation (BBR) in the evolution of the vibrational distribution of the $^{87}$Rb$^{40}$Ca$^+$ ions. As our study relies on novel molecular data compared to paper I, namely the calculation of the molecular spin-orbit couplings (SOC) using a quasidiabatic approach \cite{szczepkowski2022}, we assess their validity by computing the rate for SOC-induced non-radiative charge exchange (NRCE), which has been found to be the dominant process (in particular over RA) \cite{tacconi2011,belyaev2012} in the experiment of Ref. \cite{hall2011}, by comparison with radiative charge exchange (RCE) and RA. Our work complements the recent theoretical investigation of Ref. \cite{zrafi2020}.

The paper is organized as follows. We first present the molecular data in Section \ref{sec:data}, including the main features of the quasidiabatic approach for SOC. Then in Section \ref{sec:ra-bbr} we recall the expected vibrational population of the RbCa$^+$ molecular ions created by radiative association, and investigate its possible redistribution induced by black-body radiation. Then the main result of this paper is reported in Section \ref{sec:pd}, where we show that the created molecular ions are protected against photodissociation in the experiment. Finally Section \ref{sec:nrce} revisits the non-radiative charge exchange, with the occurrence  of a phase-matching condition which significantly influences the description of the process.

\section{Potential energy curves, electric dipole moments, and spin-orbit couplings}
\label{sec:data}

The potential energy curves (PECs), transition electric dipole moments (TEDMs), and permanent electric dipole moments (PEDMs) for numerous electronic states are calculated following the methodology and parameters described in Ref. \cite{guerout2010,aymar2012}, and already used in paper I. In a few words, it consists in using the CIPSI (Configuration Interaction by Perturbation of a Multiconfiguration wave function Selected Iteratively) program \cite{huron1973}. The diatomic molecular ion is modeled as an effective two-valence-electron system with effective core potentials (ECP) from Ref. \cite{durand1974}, complemented with core-polarization potentials (CPP) \cite{schmidt-mink1984} parametrized according to the angular momentum of the valence electron as in Ref. \cite{foucrault1992}. Two large Gaussian basis sets are employed, centered on each ionic core (namely, Rb$^+$ and Ca$^{2+}$ here). A full configuration interaction (FCI) is then achieved to obtain from the same calculation the properties of the electronic ground state and of many excited electronic states of various symmetries with the same numerical accuracy. More details are provided in the Appendix.

The energies of the eight lowest dissociation limits of RbCa$^+$  are listed in the Appendix, for the sake of clarity. By construction, our values are found to be very close to the experimental values for the Rb($5s\,^2S$)+Ca$^+$($nl\,^2L$) limits, as the parametrization of the ECPs and CPPs is performed on the atomic energy spectrum of Rb and Ca$^+$. Inherently, the discrepancy for the Rb$^+(^1S)$+Ca($4s^2\,1S$) limit can be larger than 100~cm$^{-1}$ as the atomic energies of the neutral calcium atoms result from the FCI. Our results are in agreement with those of the recent investigation of \cite{zrafi2020} based on a very similar approach, with however significant differences in the used parametrization. We note that for several asymptotes, our results are improved compared to \cite{zrafi2020}.

\begin{figure}[!ht]
\includegraphics[width=0.6\textwidth]{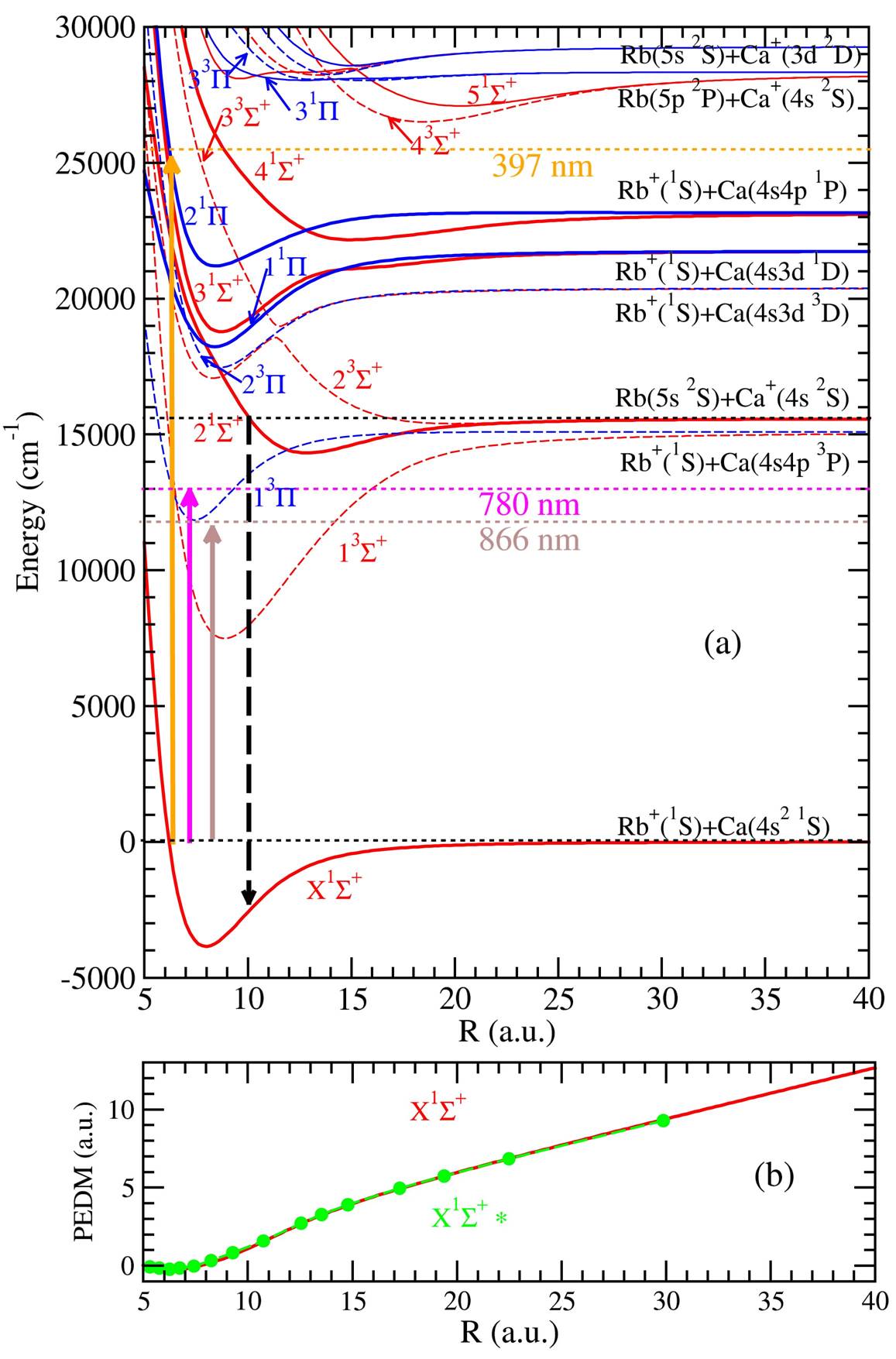}
\caption{(a) Hund$'$s case $a$ RbCa$^+$ PECs as functions of internuclear distance $R$. The origin of energies is taken at the asymptote of the electronic ground state Rb$^+(^1S)$+Ca($4s^2\,^1S$). The length of the vertical solid upward arrows represent the energies of the lasers present in the experiment \cite{hall2011} with wavelengths 397~nm (orange arrow), 780~nm (magenta arrow) and 866~nm (brown arrow). The black dashed downward arrow holds for the radiative association process (see the main text) resulting in the formation of RbCa$^+$ ions. (b) PEDM of the $X^1\Sigma^+$ state (solid red line) computed with the origin of distance taken at the RbCa$^+$ center-of-mass, compared to the theoretical results of \cite{zrafi2020} (closed green dots).}
\label{fig:PEC}
\end{figure}

Figure \ref{fig:PEC} displays the computed PECs for the $^{1}\Sigma^+$,$^{3}\Sigma^+$, $^{1}\Pi$ and $^{3}\Pi$ electronic states up to the 8$^{th}$ dissociation limit. The corresponding spectroscopic constants are reported in the Appendix, showing an overall satisfactory agreement with \cite{zrafi2020}; a few differences are visible however, which are discussed in the Appendix. There are two noticeable patterns in Figure \ref{fig:PEC}. Firstly, the 2$^1\Sigma^+$ PEC correlated to Rb($5s\,^2S$)+Ca$^+$($4s\,^2S$) (which is the entrance scattering channel in the experiment \cite{hall2011}) crosses the 1$^3\Pi$ PEC dissociating into Rb$^+(^1S)$+Ca($4s4p\,^3P$) in two locations: $R_1=11.9$~a.u. and $R_2=17.4$~a.u. (1~a.u. $\equiv a_0 = 0.0529177210903$~nm). Then non-radiative charge exchange (NRCE) can occur, induced by spin-orbit coupling (SOC), as described in \cite{tacconi2011}. Secondly, the 6$^1\Sigma^+$ and 5$^3\Sigma^+$ PECs correlated to the Rb($5s\,^2S$)+Ca$^+$($3d\,^2D$) limit, which is the other possible entrance channel in the experiment as the metastable $3d\,^2D$ Ca$^+$ level is involved in the laser cooling cycle, exhibit an avoided crossing with the 5$^1\Sigma^+$ and 4$^3\Sigma^+$ PECs respectively, in the 13$a_0$-15$a_0$ range of internuclear distances (see the magnified figure in the Appendix). Therefore, excitation exchange toward the Rb($5p\,^2P$)+Ca$^+$($4s\,^2S$) outgoing channel is expected, which will produce energetic Ca$^+(4s\,^2S)$ ions accompanied by radiative decay of the excited Rb($5p\,^2P$) atom. For completeness, figures for selected PEDMs and TEDMs are reported in the Appendix, showing a satisfactory agreement with \cite{zrafi2020}. 

In order to model the NRCE process, an accurate description of the $R$-dependent SOCs  between the 2$^1\Sigma^+$ and the 1$^3\Pi$ PECs is required, which results into their separation by an amount as small as 2~cm$^{-1}$ at their crossing point around 17$a_0$, according to \cite{tacconi2011}. Here we used an alternative method to derive this coupling, and thus further assess the dependency of the NRCE cross section and rate with respect to the SOC amplitude (see Section \ref{sec:nrce}). The central idea is to involve only the atomic spin-orbit operators $H_{SO}^{\textrm{i}} = \alpha_{SO}^{\textrm{i}} \vec{\ell}_i \cdot \vec{s}_i$ and coupling constants $\alpha_{SO}^{\textrm{i}}$ ($i=$A, B) for each of the A and B atom or ion. The resulting matrix elements are easily evaluated when the particles are at large distances considering a linear combination of atomic orbitals (LCAO) for the AB pair modeled as an effective two-electron system. 

\begin{figure}[!t]
\includegraphics[width=0.6\textwidth]{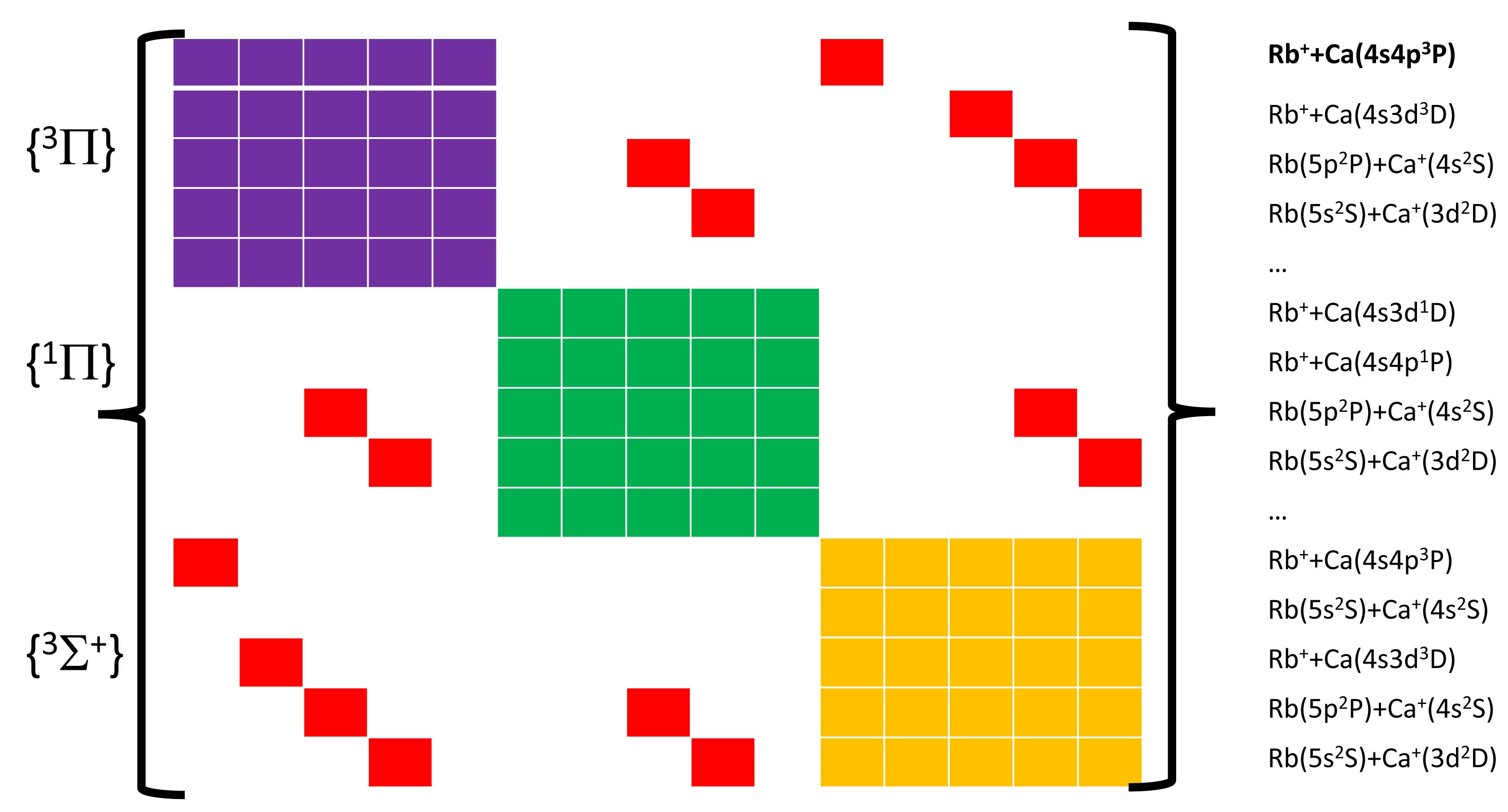}
\caption{A schematic representation of the effective Hamiltonian matrix $\textbf{H}_{\textrm{eff}}$, illustrated for the case of $\Omega=1$. Three diagonal blocks (in purple, green, and orange) are present for the three relevant Hund's case \textit{a} symmetries $^3\Pi$, $^1\Pi$, and $^3\Sigma^+$, respectively,  while the atomic SOCs (red squares) are connecting states from different blocks correlated to the same dissociation limit (thus the same fine structure manifold).}
\label{fig:Heff}
\end{figure}

Our approach is based on the \textit{quasidiabatic} method first presented in \cite{cimiraglia1985}, which, to our knowledge, has been used only once for the NaCd van der Waals molecule \cite{angeli1996}. It is described in detail in \cite{szczepkowski2022}. Briefly, we start with the Hund's case $a$ diagonal Hamiltonian matrix $\hat{\textbf{H}}^0$, which contains the adiabatic Hund's case $a$ PECs of Fig. \ref{fig:PEC} for each molecular symmetry $^{2S+1}\Lambda^{p}$, where $S$, $\Lambda$, $p$ are the quantum numbers for the total electronic spin, the projection of the total electronic angular momentum on the molecular axis, and the parity for the symmetry with respect to a plane containing the molecular axis. Considering the $N_{\textrm{ref}}$ eigenvectors associated to the $N_{\textrm{ref}}$ lowest PECs for each symmetry at a large distance $R_{\textrm{ref}}$ as a basis set representing the separated atoms, we apply a transformation, at every distance $R<R_{\textrm{ref}}$, to the corresponding $N_{\textrm{ref}}\times N_{\textrm{ref}}$ diagonal matrix. It yields a quasidiabatic non-diagonal Hamiltonian matrix $\textbf{H}_{\textrm{diab}}(^{2S+1}\Lambda^{p};R)$ which describes the initial adiabatic states in the basis set of the separated atoms. As depicted in Fig. \ref{fig:Heff}, we assemble the $\textbf{H}_{\textrm{diab}}(^{2S+1}\Lambda^{p};R)$ matrices for each symmetry involved in the spin-orbit interaction ($^3\Pi$, $^1\Pi$, $^3\Sigma^+$, for the Hund's case $c$ symmetry $\Omega=1$) into a single block-diagonal matrix $\textbf{H}_{\textrm{eff}}(\Omega^p;R)$. The blocks are connected through the atomic spin-orbit terms relevant for every atomic state (listed in Fig. \ref{fig:Heff}), according the matrices displayed in the Appendix. Then two options are available:
\begin{itemize}
    \item Diagonalizing $\textbf{H}_{\textrm{eff}}(\Omega^p;R)$ to obtain the adiabatic Hund's case $c$ PECs for every $\Omega^p$ symmetry.
    \item Performing the inverse transformation on $\textbf{H}_{\textrm{eff}}(\Omega^p;R)$ back to the original adiabatic basis, leading to a matrix $\mathcal{H}_0(R)(\Omega^p;R)$ with diagonal and off-diagonal matrix elements for the molecular spin-orbit couplings between the involved molecular symmetries. 
\end{itemize} 

Following Ref.\cite{szczepkowski2022}, we selected $N_{\textrm{ref}}=40$ Hund's case \textit{a} states in each molecular symmetry, and we optimized the transformation in order to have the quasidiabatic matrix elements converged for the eight lowest reference states.

\begin{figure}[ht]
\includegraphics[width=0.58\textwidth]{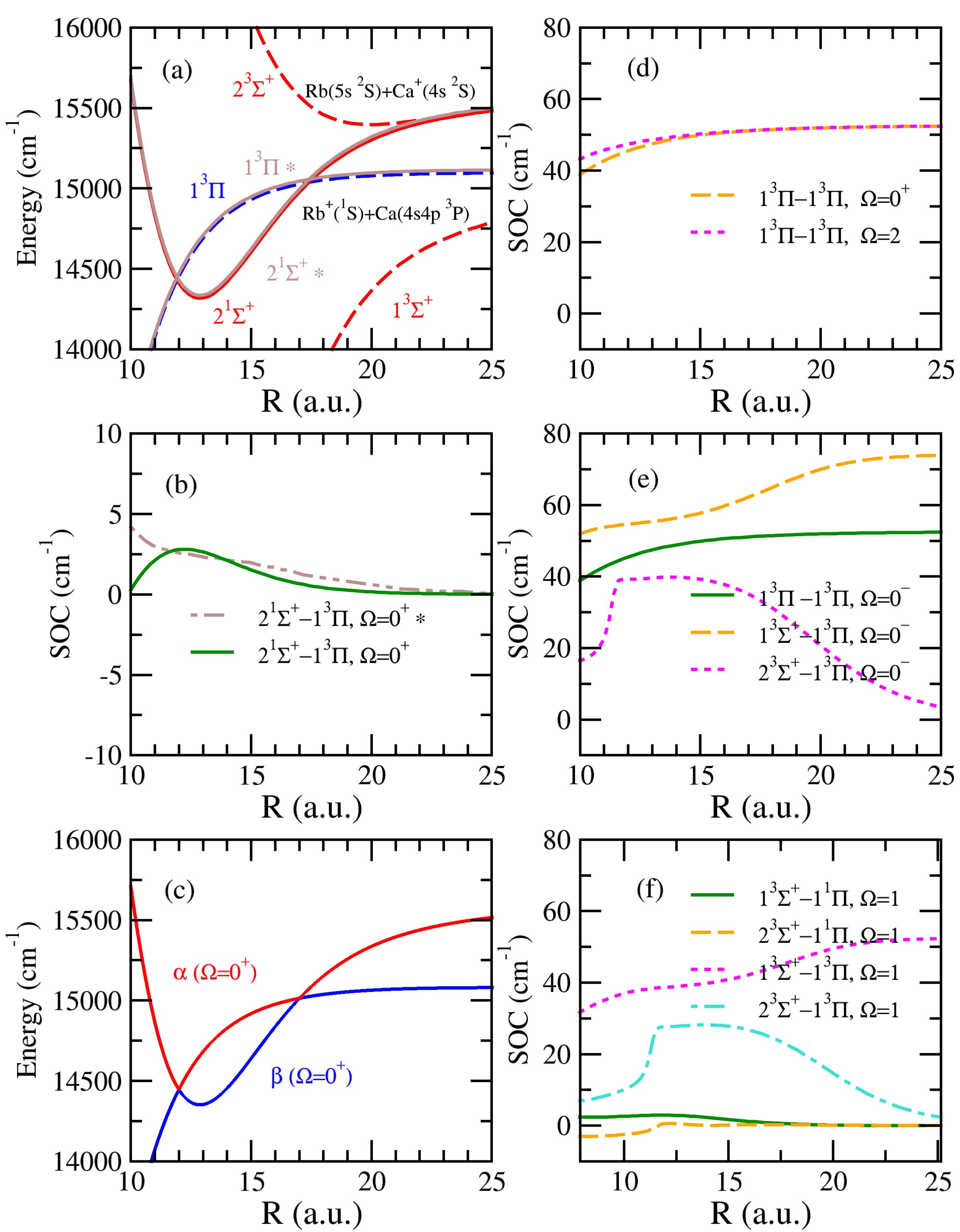}
\caption{(a) Magnified view of the RbCa$^+$ PECs involved in the NRCE process Rb($5s\,^2S$)+Ca$^+$($4s\,^2S$) $\rightarrow$ Rb$^+$+Ca($4s4p\,^3P$). The theoretical results from \cite{tacconi2011} are labeled by a star $*$ (brown line). (b) SOC between the 2$^1\Sigma^+$ and the 1$^3\Pi$ states computed in the present work for the $\Omega=0^+$ symmetry, together with the one computed in \cite{tacconi2011} (labeled with a star $*$, dashed brown line). (c) Adiabatic PECs resulting from the diagonalization of the $\textbf{H}_{\textrm{eff}}(\Omega^p;R)$ matrix (see the main text).(d) Diagonal SOC  for the $\Omega=0^+, 2$ symmetries. (e) SOCs for the $\Omega=0^-$ symmetry. (f) SOCs for $\Omega=1$ symmetry. Note that the kinks in the coupling between 2$^3\Sigma^+$ and 1$^3\Pi$ visible in the (e) and (f) panels are due to the avoided crossing around 12~a.u. of the 2$^3\Sigma^+$ and 3$^3\Sigma^+$ PECs (see Fig.\ref{fig:PEC}.)}
\label{fig:SOC}
\end{figure}

Figure \ref{fig:SOC} displays the Hund's case a PECs (panel (a)) relevant for the NRCE process Rb($5s\,^2S$)+Ca$^+$($4s\,^2S$) $\rightarrow$ Rb$^+$+Ca($4s4p\,^3P$), the associated SOCs extracted from the diagonal and off-diagonal matrix elements of $\mathcal{H}_0(R)(\Omega^p;R)$ (panels (b), (d)-(f)), and the eigenvalues of $\textbf{H}_{\textrm{eff}}(\Omega^p;R)$ (panel (c)). The Rb$^+$+Ca($4s4p\,^3P$) asymptote is now split into three components Rb$^+$+Ca($4s4p\,^3P_{0,1,2}$), and the weakly $R$-dependent SOCs between the corresponding molecular states reflect the magnitude of the SO splitting of atomic Ca($4s4p\,^3P$). As their PECS are correlated to different asymptotes, the SOC $W^{SO}(R)$ between the $^1\Sigma^+$ and $^3\Pi$ crossing states (panel (b), $\Omega=0^+$), is weak at their crossing points ($W^{SO}(R_1=11.9 a_0)=2.77$~cm$^{-1}$, and $W_{SO}(R_2=17.4 a_0)=0.52$~cm$^{-1}$. This result is in good agreement with the published determination from Ref.\cite{tacconi2011} in this region, as well as the PECs themselves (panel (a)).

\section{Radiative association of R\lowercase{b}C\lowercase{a}$^+$ and black-body radiation}
\label{sec:ra-bbr}

The RA process was studied in paper I for a series of ionic molecular species composed of Rb and alkaline-earth ions, including RbCa$^+$. It is illustrated by the dashed black arrow in Fig.\ref{fig:PEC}, namely the spontaneous emission of a photon during the collision between cold ground-state Rb atoms and Ca$^+$ ions at a relative energy $E_{\textrm{coll}}$ via the $2^1\Sigma^+$ state toward the $X^1\Sigma^+$ state with a typical (constant) rate coefficient of about $10^{-5}$ the Langevin rate, namely $\simeq 3 \times 10^{-14}$~cm$^3$s$^{-1}$. Assuming an averaged density of Rb atoms in the MOT of $10^9$~cm$^{-3}$ \cite{hall2011}, the RA rate has a magnitude of about $10^{-5}$~s$^{-1}$. 

The population distribution over the rovibrational levels $|v'_f,J'_f\rangle$ of the RbCa$^+$ $X^1\Sigma^+$ ground state (the $f$ channel), starting from the entrance channel $i$ with a collision energy $E_{\textrm{coll}}$ and a rotational state (or partial wave) $J_i$ (thus $J_f=J_i\pm 1$), is described as the fraction $F(v'_f,J'_f,E_{\textrm{coll}},J_i)$ of the total population of the RbCa$^+$ bound levels induced by RA.
\begin{equation}
\begin{aligned}
F(v'_f,J'_f,E_{\textrm{coll}},J_i)=\dfrac{\omega_{v'_f,J'_f,E_{\textrm{coll}},J_i}|<X^1\Sigma^+,v'_f,J'_f|D_{fi}(R)|2^1\Sigma^+,E_{\textrm{coll}},J_i>|^2}{\sum_{v'_f=0}^{v'_{max}}\omega_{v'_f,J'_f,E_{\textrm{coll}},J_i}|<X^1\Sigma^+,v'_f,J'_f|D_{fi}(R)|2^1\Sigma^+,E_{\textrm{coll}},J_i>|^2},
\end{aligned}
\label{eq:fraction}
\end{equation}
where the the energy difference between the corresponding two levels is $\hbar \omega_{v'_f,J'_f,E_{\textrm{coll}},J_i}$. The $R$-dependent TEDM $D_{fi}(R)$ between the $X^1\Sigma^+$ and the $2^1\Sigma^+$ states was given in paper I. The ket $|E_{\textrm{coll}},J_i>$ represents the energy-normalized radial continuum wave function which describes the two colliding particles in the $2^1\Sigma^+$ entrance channel. The summation is performed up to the uppermost calculated $X^1\Sigma^+$ vibrational level $v_f^{\textrm{max}}=141$. 

\begin{figure}[t]
\centering
\includegraphics[width=0.55\textwidth]{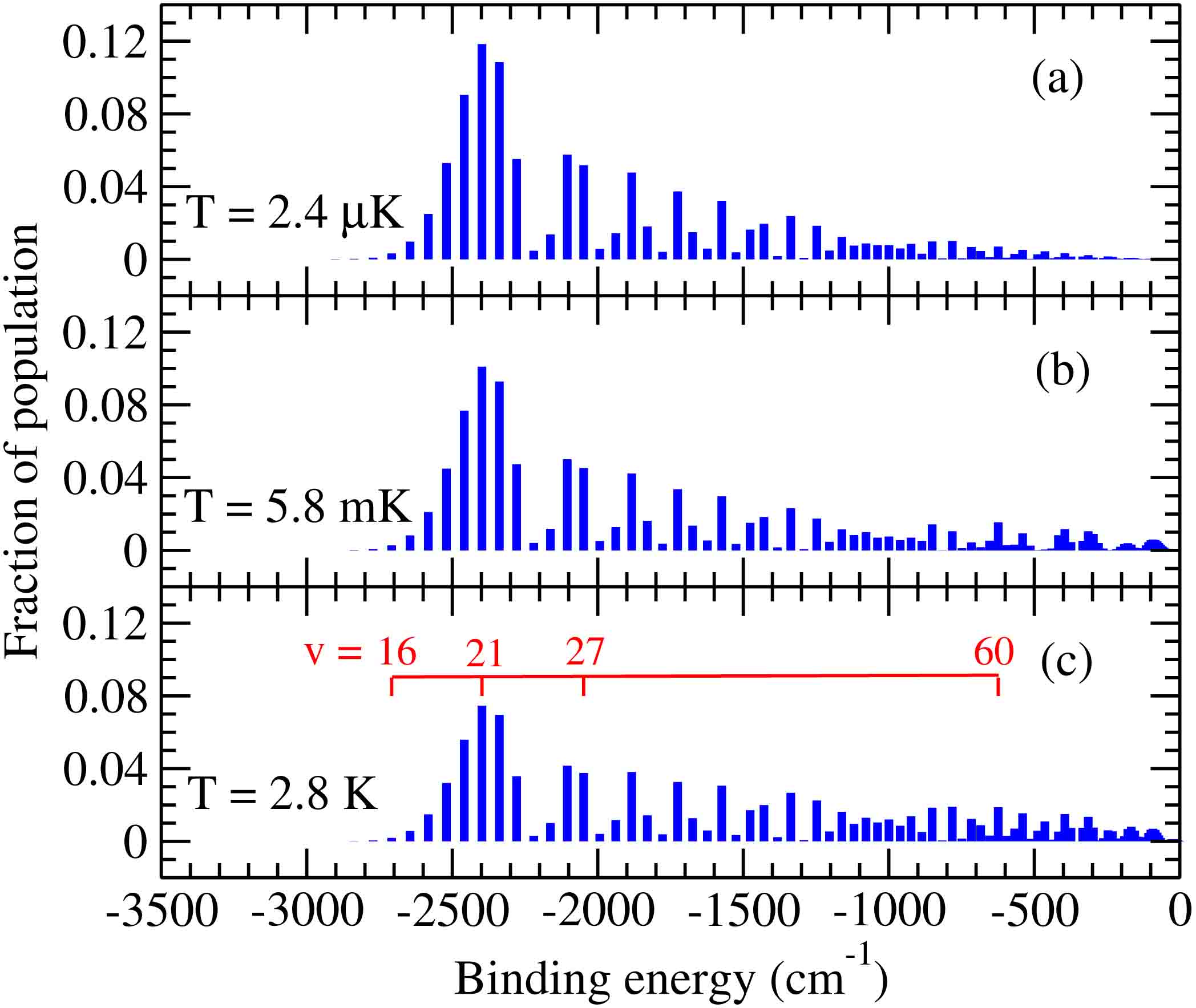}
\caption{Fraction of population (by RA  from the 2$^1\Sigma^+$ entrance channel) of the RbCa$^+$ ground-state vibrational levels (displayed in red in panel (c)) as a function of their binding energy, for three selected collision energies  (a) $\epsilon_i/k_B=$ 2.4~$\mu$K, (b) $\epsilon_i/k_B=$5.8~mK, and (c) $\epsilon_i/k_B=$2.8~K. }
\label{fig:RA}
\end{figure}

In the experiment \cite{hall2011,hall2013a}, the kinetic energy of the collision is determined by that of the Ca$^+$ ion, which could span a range of equivalent temperatures up to a few Kelvin. In Figure \ref{fig:RA}, we display the distribution of Eq.\ref{eq:fraction} for $J_i=0$, $J_f=1$ and for different values of colliding energies $\epsilon_i$ corresponding to  temperatures $E_{\textrm{coll}}/k_B=$ 2.4~$\mu$K, 5.8~mK, and 2.8~K. The distribution is independent of the temperature and obviously consistent with the results of paper I at 100~$\mu$K, and those of \cite{zrafi2020} at 1~$\mu$K. In particular, we find that the RbCa$^+$ molecular ions are preferably formed in deeply-bound vibrational levels with a maximum fraction centered at $v_f=21$. More importantly, $96\%$ of the formed  molecular ions are mainly distributed in $v_f=16-60$, and the deeper bound levels $v_f\le 15$ are unlikely to be populated by RA.

Another feature of the experiment is that the formed ground-state RbCa$^+$ ions reside quite a long time ($\approx$ minutes) in the trap, so that in principle their vibrational population could be rearranged due to the influence of black-body radiation (BBR) at room temperature (say, 300~K). The BBR energy density $\rho(\omega)$ for a given frequency $\omega$ in atomic units ($\hbar=e=m_e=1$)  (see for instance \cite{vanhaecke2007a}) is given by
\begin{equation}
\begin{aligned}
\rho(\omega)=\dfrac{2 \omega^3}{ \pi c^3} N(\omega) \equiv \dfrac{2 \omega^3}{ \pi c^3} \dfrac{1}{exp\left(\dfrac{\omega}{k_B T}\right)-1},
\end{aligned}
\end{equation}
where $N(\omega)$ stands for the number of BBR photons. Assuming that a vibrational level $v_i$ (with binding energy $\epsilon_{v_i}$, assuming a rotational state $J_i=0$) of the RbCa$^+$ ground state is exposed to BBR, transitions towards other ground state vibrational levels $v_f$ (with binding energy $\epsilon_{v_f}$, assuming $J_f=0$) can occur. The inverse of the resulting lifetime $\tau_{v_i}$ of the $v_i$ level can be expressed as
\begin{equation}
\begin{aligned}
\tau^{-1}_{v_i}=\sum_{\epsilon_{v_f}< \epsilon_{v_i}}\Gamma^{SpE}+\sum_{\epsilon_{v_f}< \epsilon_{v_i}}\Gamma^{StE}+\sum_{\epsilon_{v_i}< \epsilon_{v_f}}\Gamma^{Abs},
\end{aligned}
\label{eq:bbr1}
\end{equation}
where $\Gamma^{SpE}$, $\Gamma^{StE}$, and $\Gamma^{Abs}$ are the rates for spontaneous emission, stimulated emission, and absorption, respectively. Defining $\omega_{fi} = |\epsilon_{v_f} - \epsilon_{v_i}|$, we have
\begin{equation}
\begin{aligned}
\Gamma^{SpE}=\dfrac{4{\omega_{fi}}^3}{3c^3}|<v_f,J_f=1|D_X(R)|v_i,J_i=0>|^2 ,
\end{aligned}
\label{eq:bbr2}
\end{equation}
and 
\begin{equation}
\begin{aligned}
\Gamma^{StE} \equiv \Gamma^{Abs} =\Gamma^{SpE}N(\omega).
\end{aligned}
\label{eq:bbr3}
\end{equation}
where $D_X(R)$ is the $R$-dependent permanent electric dipole moment in the ground state of RbCa$^+$ (Fig. \ref{fig:PEC}b).

\begin{figure}
\centering
\includegraphics[width=0.75\textwidth]{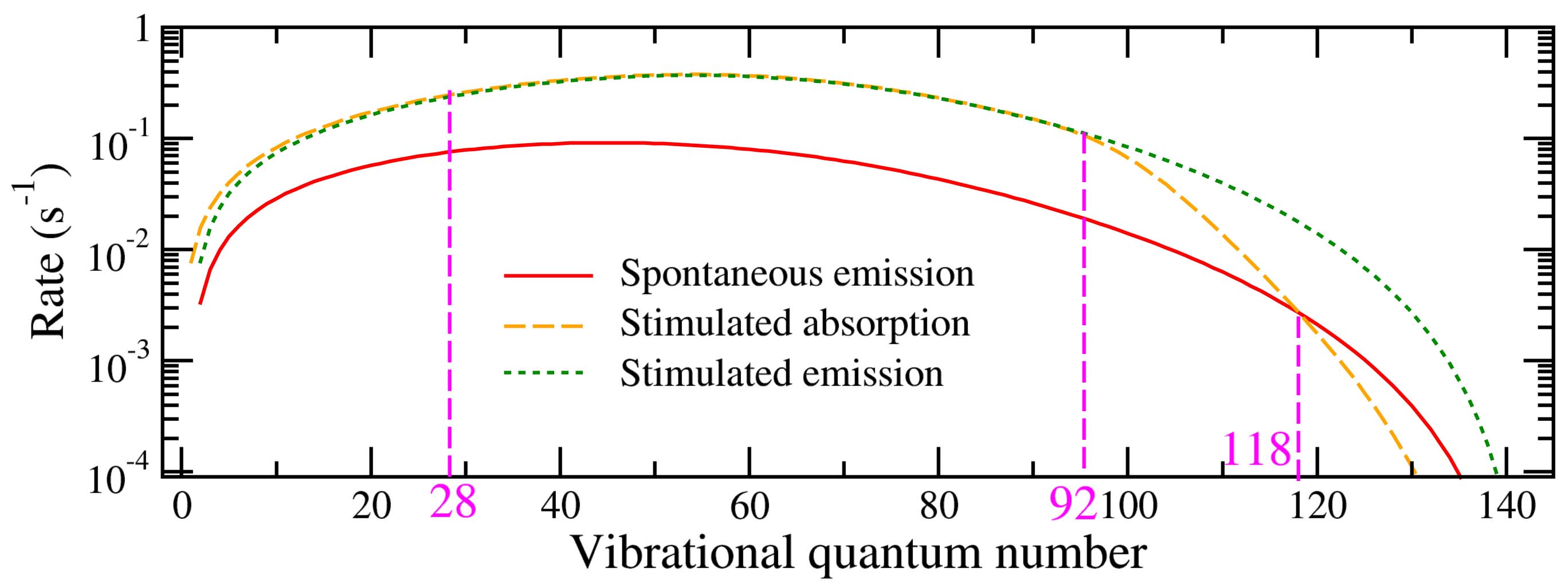}
\caption{Rates for spontaneous/stimulated emission and absorption of the vibrational levels of the RbCa$^+$ $X^1\Sigma^+$ ground state. The stimulated radiation rate is the sum of the absorption and stimulated emission rates. The vertical dashed line at $v=28$ marks the photodissociation threshold for the 397~nm laser (see section \ref{sec:pd}). At $v=92$, the absorption rate becomes smaller than the stimulated emission rate, and than the spontaneous emission rate at $v=118$. }
\label{fig:BBR}
\end{figure}

\begin{figure}
\centering
\includegraphics[width=0.6\textwidth]{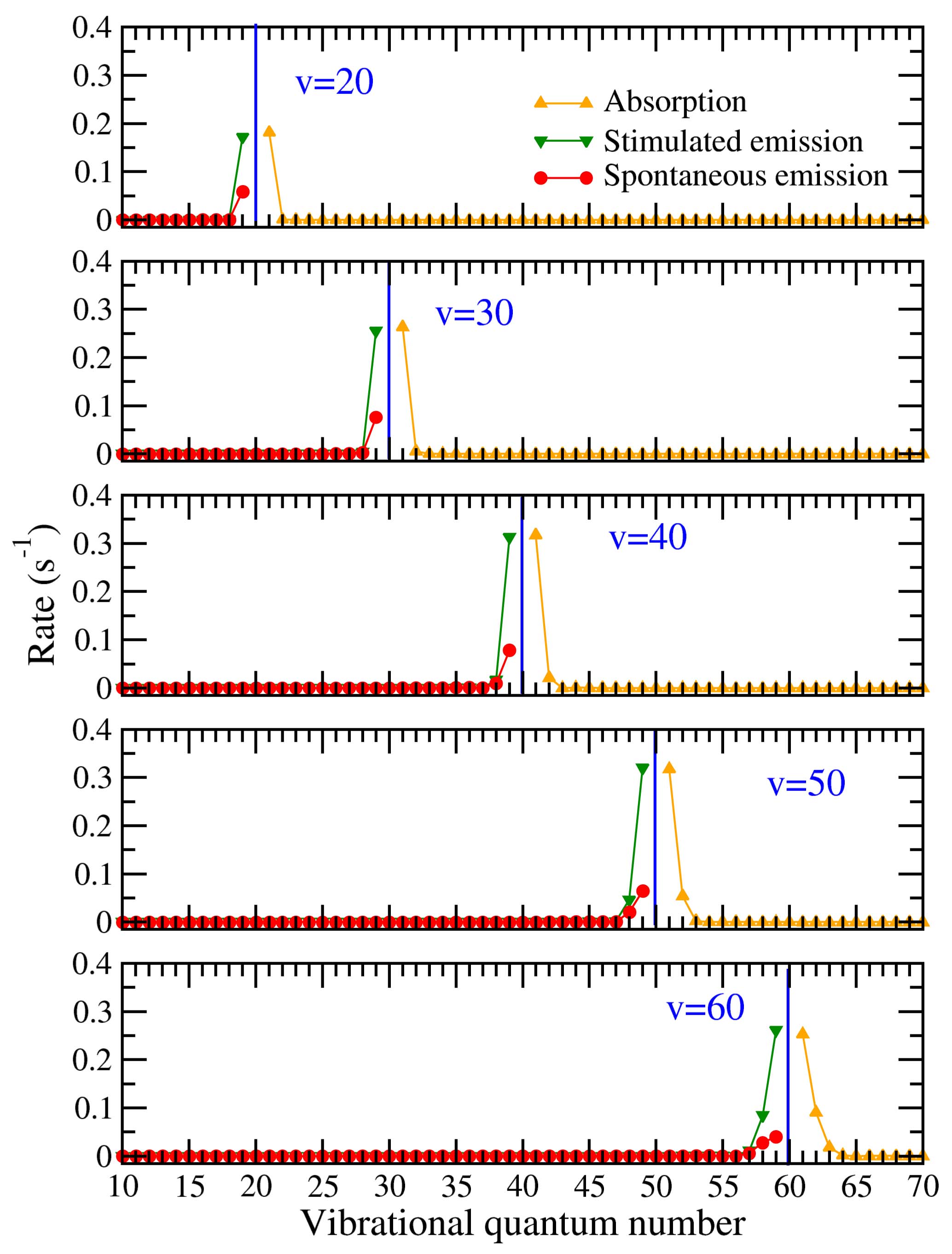}
\caption{Rates for spontaneous/stimulated emission and absorption for the $X^1\Sigma^+$ ground state vibrational levels $v_i=20,30,40,50, 60$ towards every $X^1\Sigma^+$ final level $v_f$ for $J_i=0 \to J_f=1$.}
\label{fig:vrate}
\end{figure}

The calculated rates for spontaneous emission, stimulated emission, and absorption  are displayed in Fig. \ref{fig:BBR}. The former process is significantly less efficient than the stimulated ones but, by comparison, are larger than the RA rate. In other words, once a molecular ion is created by RA in the range of vibrational levels [16-60] (Fig. \ref{fig:RA}), it will be sensitive to this class of processes. However the modification of the vibrational distribution is likely to be minimal, as suggested by Fig. \ref{fig:vrate}: the transitions induced by BBR almost exclusively occur toward the next vibrational level (\textit{i.e.} $v_f-v_i=\pm 1$), while the absorption and the stimulated emission rates are very similar to each other.

\section{Photodissociation of R\lowercase{b}C\lowercase{a}$^+$ created by radiative association}
\label{sec:pd}

One of the main feature of the hybrid-trap experiments is that several lasers are involved and switched on during the measurements. Therefore, the created molecular ions are likely to be exposed to strong photodissociation (PD), as previously emphasized for Rb$_2^+$ ions \cite{jyothi2016} and RbBa$^+$ ions \cite{mohammadi2021}, while in somewhat different experimental conditions. In the experiment of \cite{hall2011,hall2013a} three cooling lasers are present (Fig. \ref{fig:PEC}) with the following wavelengths: 397~nm and 866~nm for the $4s\,^2S_{1/2} \to 4p\,^2P_{1/2}$  cooling transition and the $3d\,^2D_{3/2} \to 4p\,^2P_{1/2}$ repumping transition in Ca$^+$, respectively, and 780~nm for $5s\,^2S_{1/2} \to 5p\,^2P_{3/2}$ cooling transition in $^{87}$Rb. We see in Fig.~\ref{fig:PEC} that only the laser at 397~nm is expected to photodissociate the ground-state RbCa$^+$ ions. Indeed, due to the selection rules for dipolar transitions, only dissociation channels associated with the continuum region of $^1\Sigma^+$ and $^1\Pi$ states can be reached from the ground electronic state. The energies of 780~nm and 866 nm photons are too low to induce photodissociation, even from a weakly bound, i.e., vibrationally highly excited, RbCa$^+$ molecular ion in its electronic ground state. The 397~nm laser is the only one which can dissociate  ground-state RbCa$^+$ leading to five possible exit channels.

Our analysis is performed considering Hund's case $a$ PECs, neglecting SOCs which are rather small compared to the photon energies, so that they should not significantly affect the results. We calculate the PD rate of a RbCa$^+$ ground state molecular ion created in a rovibrational level $v_i,J_i$, considering all the excited molecular states $f$ that are energetically open with the above lasers. The final continuum state is labeled by $\epsilon_f,J_f$, where $\epsilon_f$ is the relative kinetic energy of the dissociated atom-ion pair above the dissociation energy of the $f$ state. The PD rate is given by
\begin{equation}
\begin{aligned}
\Gamma_{\textrm{PD}} (v_i,J_i,\epsilon_f,J_f)=\sigma_{\textrm{PD}}(v_i,J_i,\epsilon_f,J_f)K=S_{J_i,J_f}\dfrac {4 \pi^2a_0^2}{3\hbar c} E_{\lambda}| <\epsilon_f,J_f|D_{fi}(R)|v_i,J_i> |^2\dfrac{I}{\hbar \omega_{\lambda}},
\end{aligned}
\end{equation}

where $\sigma_{PD}$ is the state-to-state dissociation cross section, $K$ is the flux of dissociating light with intensity $I$, and $E_{\lambda}=hc/\lambda\equiv \hbar \omega_{\lambda}$ is the photon energy with wavelength $\lambda$ and circular frequency $\omega$. The final energy $\epsilon_f=E_{\lambda} - |E_{i}-E_{f}|-|E_{v_i,J_i}|$ is obtained from the difference of energies of the dissociation limits of the initial ($E_{i}$) and final ($E_{f}$) channels, and from the binding energy $E_{v_i,J_i}$ of the initial level ($v_i,J_i$). The H\"onl-London factors $S_{J_i,J_f}$ are also included. We note, however, that according to \cite{mohammadi2021}, the total PD cross section is independent of $J_i$ when it is summed over the contributions of the P ($J_f=J_i-1$), Q ($J_f=J_i$, and R ($J_f=J_i+1$) branches. Therefore we assume $J_i = 0$ in the following, for the sake of simplicity.

\begin{figure}
\centering
\includegraphics[width=0.6\textwidth]{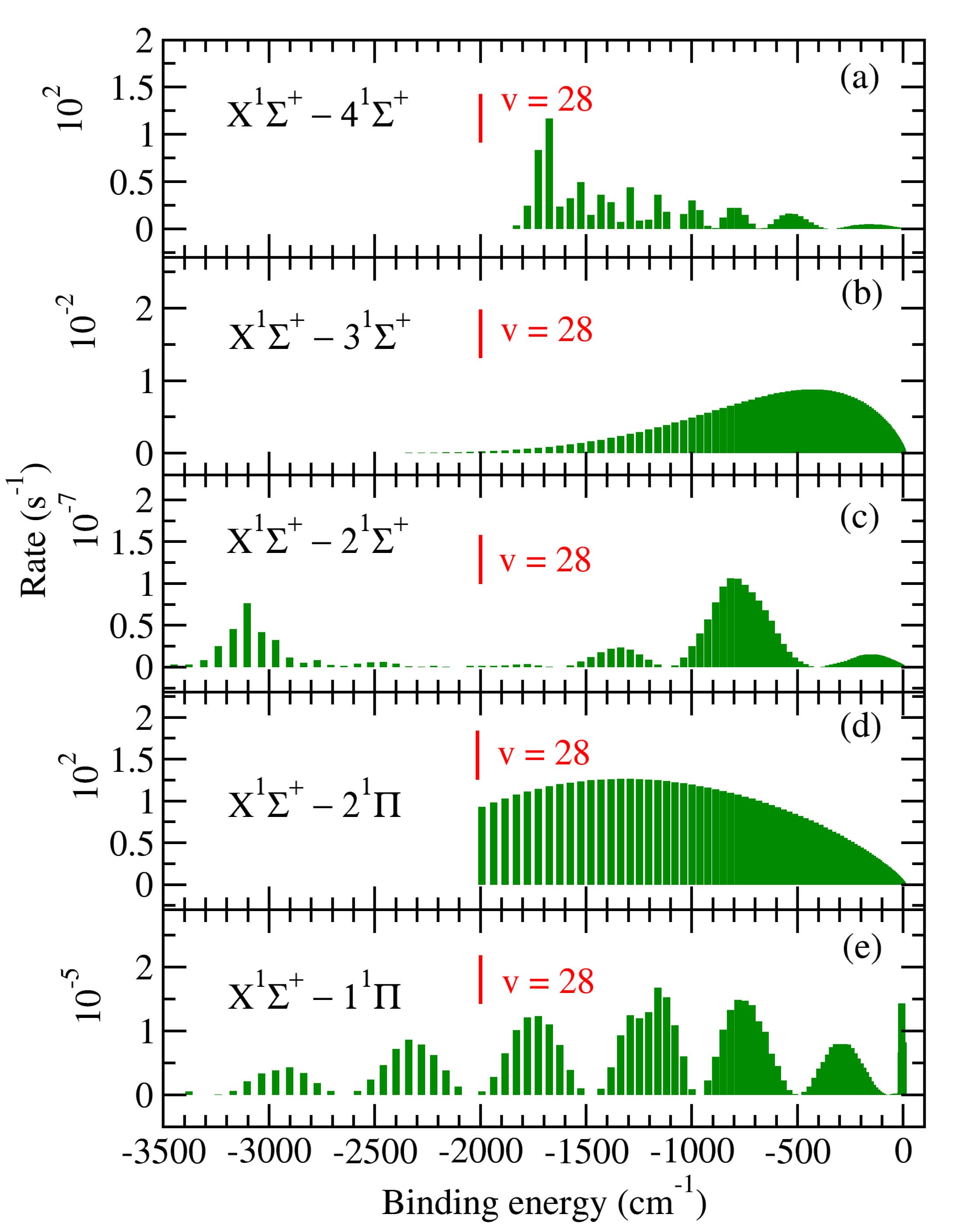}
\caption{Calculated photodissociation rates induced by the 397~nm cooling laser, assuming a laser intensity of  $I=400$~mW/cm$^2$, as functions of the binding energy $E_{v_i}$ of the $X^1\Sigma^+$ vibrational levels $v_i$, for the five open exit channels (a) 4$^1\Sigma^+$, (b) 3$^1\Sigma^+$, (c) 2$^1\Sigma^+$, (d) 2$^1\Pi$, and (e) 1$^1\Pi$. The photodissociation threshold $v_i=28$ is specified in red.}
\label{fig:PD-rate}
\end{figure}

The corresponding PD rates for a laser intensity of  $I=400$~mW/cm$^2$ \cite{hall2013a} are represented in Fig. \ref{fig:PD-rate} as a function of the binding energy $E_{v_i}$ of the initial ground-state level. Two channels, 4$^1\Sigma^+$ and 2$^1\Pi$, are reached with a significant rate compared to the others: this is expected from the matching in distance of the repulsive branch of their PECs with the extension of the ground state potential well, for the energy of the 397~nm laser. We also see that, due to the large energy difference between the ground-state PEC and the final ones, vibrational levels below $v_i = 28$ cannot be photodissociated. 
According to the definition of the final energy $\epsilon_f$ given above, only ground-state ions created in $v \ge 28$ can undergo PD with $\epsilon_f > 0$. From the vibrational populations created by RA reported in Section \ref{sec:ra-bbr}, we infer that about 41\% of the molecular ions with $v_i >28$ would be dissociated by the 397~nm laser, while about $59\%$ of the molecular ions created in levels below $v_i=28$) are protected against PD, and thus could be detected, in agreement with the experimental observations of \cite{hall2011,hall2013a}.

\section{Non-radiative charge exchange between ground state R\lowercase{b} and C\lowercase{a}$^+$}
\label{sec:nrce}

A previous quantum-scattering study of the NRCE between ground-state Ca$^+$ and Rb has already been reported in \cite{tacconi2011,belyaev2012}. As our molecular data, namely the PECs and SOCs, are noticeably different from the ones of these references, we recalculate the corresponding rates in order to evaluate their sensitivity. As recalled in Section \ref{sec:data}, the NRCE process involves the 2$^1\Sigma^+$ (entrance channel) and the 1$^3\Pi$ PECs (outgoing channel), which cross each other in two locations, $R_1 = 11.9 a_0$ and $R_2 = 17.4 a_0$ (Fig. \ref{fig:SOC}a). In Hund's case $c$, the former state gives rise to a single state with $\Omega=0^+$, and the latter to four states with $\Omega=0^+, 0^-, 1, 2$, correlated to the dissociation limits Rb$^+$($^1S$)+Ca($4s4p\,^3P_j$), with $j=0, 1, 2$. The corresponding spin-orbit Hamiltonian matrices are presented in the Appendix, and involve the $R$-dependent SOCs computed in Section \ref{sec:data} and displayed in Fig. \ref{fig:SOC}b, d, e, f.

Thus only the $\Omega=0^+$ symmetry is relevant for NRCE through the 2$\times$2 SO matrix reported in the Appendix. The SOC between the $2^1\Sigma^+$ and $1^3\Pi$ states (Fig. \ref{fig:SOC}b) vanishes at large distance as their PECs are correlated to different dissociation limits. It smoothly increases in the crossing region due to the configuration mixing as the internuclear distance $R$ decreases. The SOC diagonal term for $1^3\Pi$ (Fig. \ref{fig:SOC}d) has a finite value at large distance corresponding to the SO splitting of the Ca atom (see Appendix), and smoothly decreases with decreasing $R$ due to this configuration mixing, too. Following \cite{tacconi2011,belyaev2012}, we do not consider any rotational coupling between the $2^1\Sigma^+$ and $1^3\Pi$ states as they have different spin multiplicity, so that the rotational quantum number (or the partial wave) noted $\ell$ in this section is assumed to be conserved during the NRCE.

\begin{figure}
\centering
\includegraphics[width=0.7\textwidth]{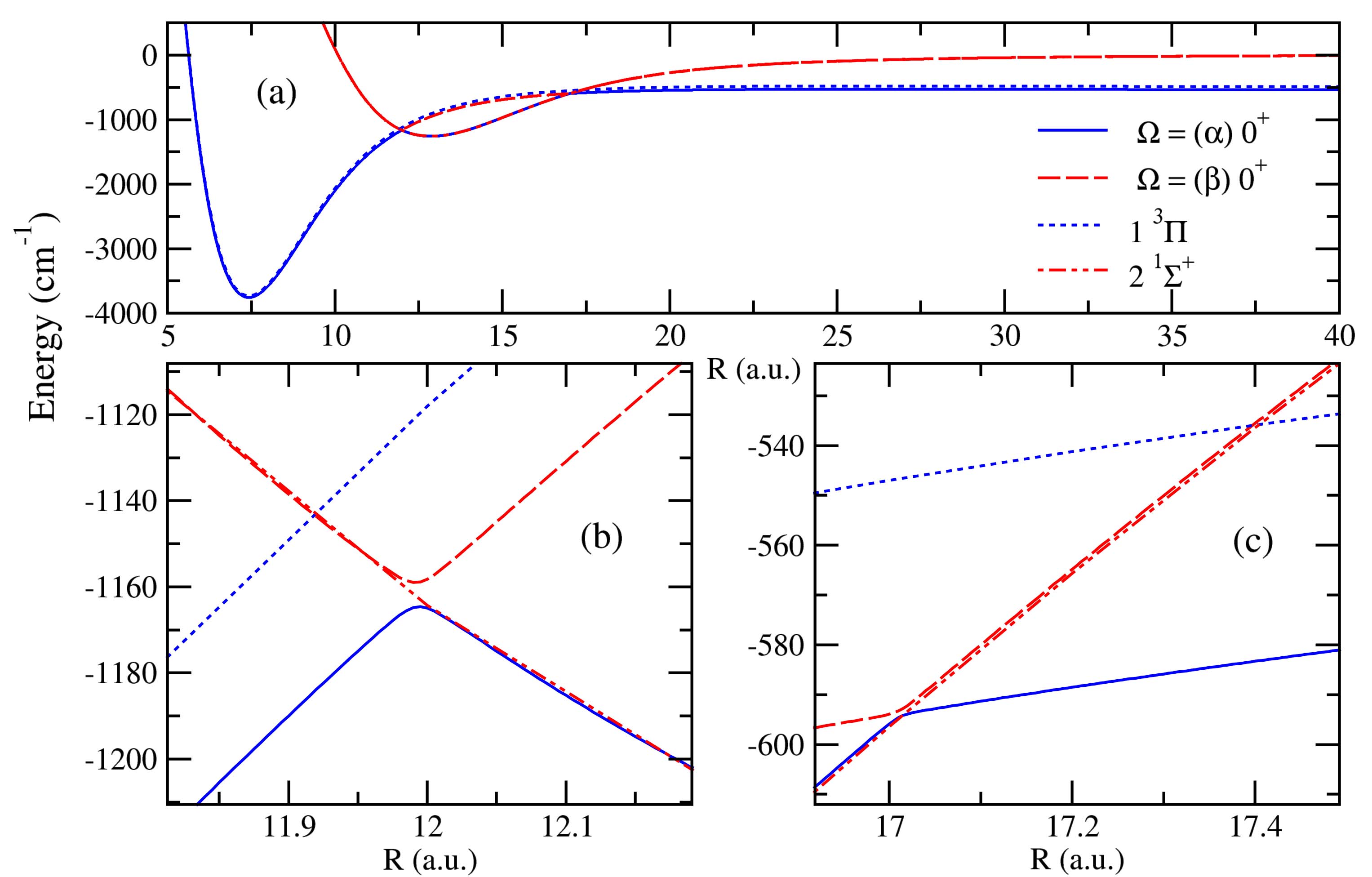}
\caption{(a) RbCa$^+$ Hund's case $a$ $2^1\Sigma^+$ and $1^3\Pi$ states, and Hund's case $c$  PECs labelled by $(\alpha)$ and $(\beta)$ for $\Omega=0^+$; (b) magnification around the crossing regions  at $R_1=11.9 a_0$, and (c) at $R_2=17.4 a_0$.}
\label{fig:0+}
\end{figure}

The $\Omega=0^+$ PECs obtained after diagonalization of the 2$\times$2 SO matrix are drawn in Fig. \ref{fig:0+}, together with a magnification of the region of the crossings. Two rather small avoided crossings are visible, slightly shifted compared to the location of the crossing between Hund's case a PECs, due to the energy shift of the $1^3\Pi$ dissociation limit equal to $A_{sp}/3 \equiv (E(^3P_{2})-E(^3P_{0}))/3$ (see the Appendix) which is induced by the diagonal term of the SOC matrix. These avoided crossing are expected to bring prominent contribution to NRCE \cite{tacconi2011,belyaev2012}.

\begin{figure}
\centering
\includegraphics[width=0.7\textwidth]{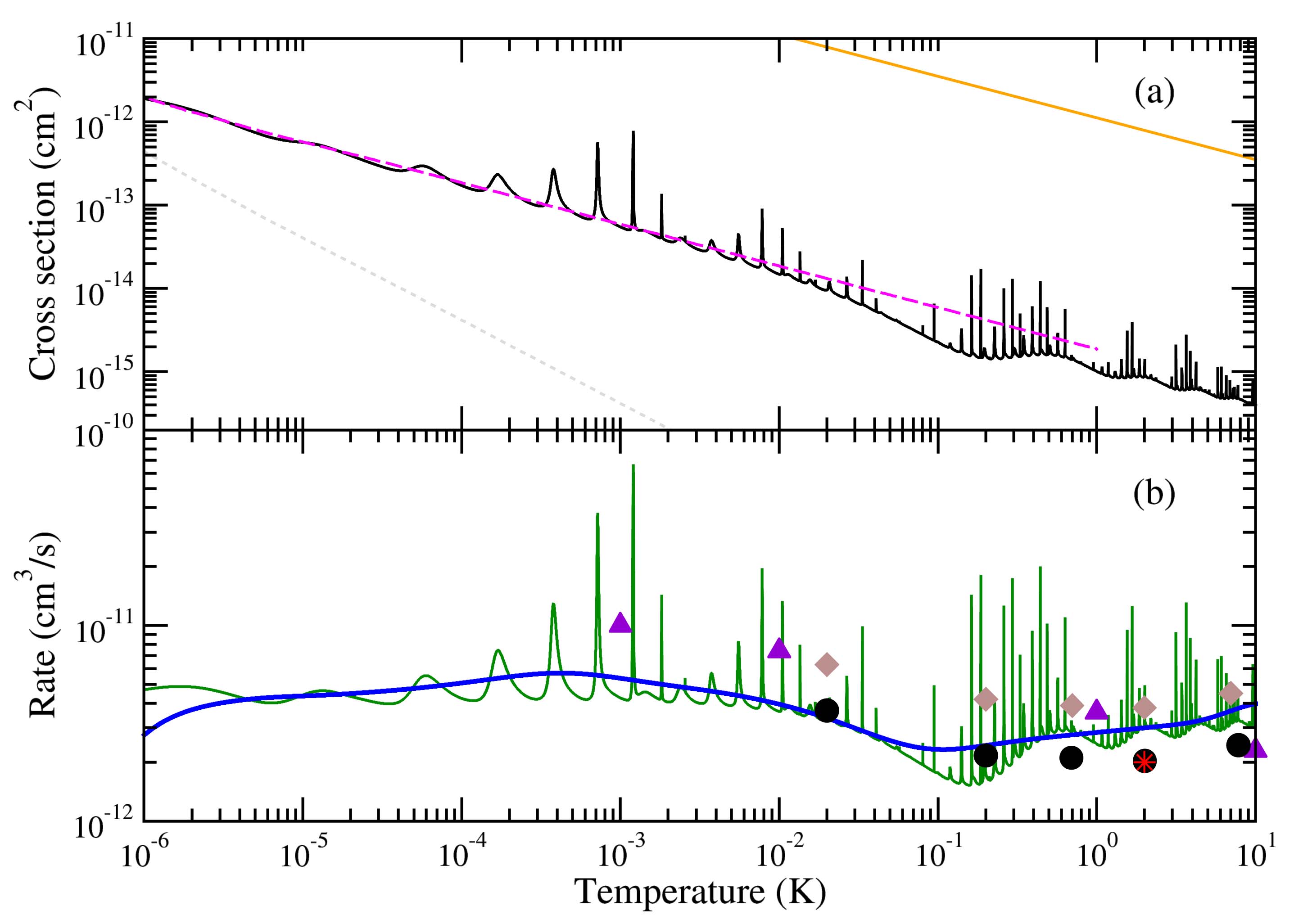}
\caption{(a) Computed cross section (solid black curve) for the NRCE process Rb($5s\,^2S$)+Ca$^+$($4s\,^2S$) $\to$ Rb$^+ (^1S)$+Ca($4s4p\,^3P$), as a function of the temperature $T=E_{\textrm{coll}}/k_B$ related to the relative collision energy $E_{\textrm{coll}}$ compared to the Langevin cross section (solid orange line) and to the expression of Eq. \ref{eq:s-signature} (first term:  dotted grey line; second term: dashed magenta line). (b)  Non-thermalized (solid green oscillatory line) and Boltzmann-thermalized (superimposed solid blue line) NRCE rate coefficients, compared to the theoretical results of \cite{tacconi2011} averaged with a Boltzmann distribution (violet triangles), with  the distribution determined by the ion velocities (brown diamonds, see Fig. 5c of \cite{hall2013b}), to the down-scaled experimental data (black circles) of \cite{hall2013a} (see text), and to the experimental value at 2~K of \cite{hall2013a} (red star).}
\label{fig:NRCEcs}
\end{figure}

To precisely account for the contribution of these two avoided crossings, we solve the set of two coupled Schr\"odinger equations for the two electronic states coupled by the SO interaction, for each $\ell$, using the standard log-derivative propagator \cite{manolopoulos1986}, yielding the usual scattering matrix $\boldsymbol{S}$. The cross section for the NRCE process is determined at each relative collision energy $E_{\textrm{coll}}=\mu v_{\textrm{coll}}^2/2$ (with $v_{\textrm{coll}}$ the relative velocity) by the off-diagonal elements $S_{if}$ of the $\boldsymbol{S}$ matrix, and expressed as a sum over partial waves $\ell$
\begin{equation}
\begin{aligned}
\sigma_{\textrm{NRCE}}(E_{\textrm{coll}})= p \dfrac{\pi \hbar^2}{2 \mu E_{\textrm{coll}}} \sum_{\ell=0}^{\infty} (2\ell+1)|S_{if}|^2,
\end{aligned}
\label{eq:Xs}
\end{equation}
where $p=1/4$ is the statistical weight for the population of the $X^1\Sigma^+$ state from the entrance channel, and $\mu$ is the reduced mass of RbCa$^+$. We find that the transition probability $|S_{if}|^2$ of the NRCE is equal to 0.006 for $\ell=0$ at low collision temperatures, which is consistent with the result of 0.007 of \cite{tacconi2011}. This is a first sign that the differences in the PECs and SOCs between our work and those of \cite{tacconi2011} do not significantly influence the probability. The NRCE cross section is converged for every energy with respect to a maximum amount of partial waves, from 5 at $T=E_{\textrm{coll}}/k_B=10^{-6}$~K, up to 122 at $T=E_{\textrm{coll}}/k_B=10$~K. The results are displayed in Fig.\ref{fig:NRCEcs}a depending on collision energy $E$. It exhibits the expected $E_{\textrm{coll}}^{-1/2}$ slope of the Langevin model, and numerous shape resonances in the entrance channel.

\begin{figure}
\centering
\includegraphics[width=0.6\textwidth]{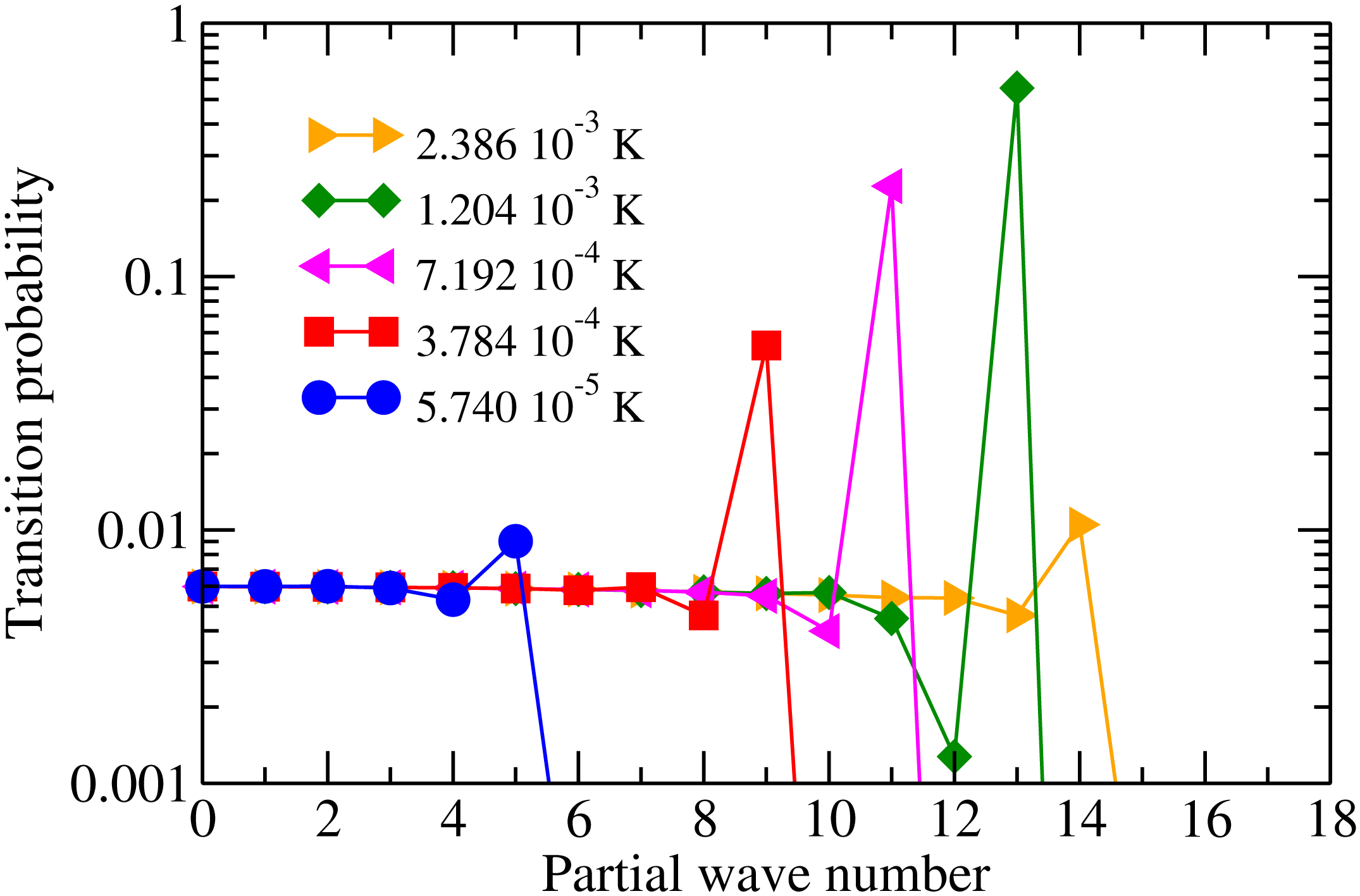}
\caption{Computed NRCE probability as a function of the partial wave $\ell$, for selected temperatures $T=E_{\textrm{coll}}/k_B$ where a shape resonance (the peak on each curve) occurs. }
\label{fig:NRCE-probas}
\end{figure}

It is worthwhile noticing that just like the resonant scattering processes (charge exchange, spin exchange) involving two interfering channels with the same asymptotic energy  \cite{li2012,sikorsky2018b,cote2018,pandey2020}, a clear manifestation of the so-called "phase-locking" is visible in the case of the present \textit{non-resonant} CE process (Fig.~\ref{fig:NRCE-probas}): the transition probability remains constant, equal to the one of the $s$-wave, for many partial waves at a given energy, and over a broad range of energies. This is due to the small energy difference between the two asymptotes Rb($5s\,^2S$)+Ca$^+$($4s\,^2S$) and Rb$^+(^1S)$+Ca($4s4p\,^3P)$ (see the Appendix) compared to the well depth of the relevant PECs (see the Appendix). Therefore this pattern is also relevant for ''quasi-resonant'' processes. Recasting Eq.\ref{eq:Xs} as
\begin{equation}
\begin{aligned}
\sigma_{\textrm{NRCE}}(E_{\textrm{coll}})=p \dfrac{\pi}{2\mu E_{\textrm{coll}}} \sum_{\ell=0}^{\infty}(2\ell+1)\sin^2(\delta_{\ell}^i-\delta_{\ell}^f),
\label{eq:phaselocking}
\end{aligned}
\end{equation}
where $\delta_{\ell}^i$ (resp. $\delta_{\ell}^f$) is the phase shift for the $i$, or $2^1\Sigma^+$ (resp.$f$, or $1^3\Pi$) channel, we see that the phase shift difference $\Delta\delta_{\ell}=\delta_{\ell}^i-\delta_{\ell}^f$ is almost $\ell$-independent, and we can determine $\sin^2\Delta\delta_{\ell}$ from our results at $\ell=0$. In \cite{cote2018}, such an approximation was named as ''s-wave signature'', and the cross section at an energy $E_{\textrm{coll}}$ is expressed as
\begin{equation}
\begin{aligned}
\sigma_{NRCE}(E_{\textrm{coll}})=p \left(\dfrac{\pi}{2\mu E_{\textrm{coll}}}+\sigma_L(E_{\textrm{coll}})\right)\sin^2\Delta\delta_0(E_{\textrm{coll}}),
\end{aligned}
\label{eq:s-signature}
\end{equation}
where $\Delta\delta_0(E_{\textrm{coll}})$ is the $s-$wave phase shift depending on the collision energy $E_{\textrm{coll}}$, and $\sigma_L(E_{\textrm{coll}})$ the Langevin cross section, which is then rescaled with the $\sin^2\Delta\delta_0(E_{\textrm{coll}})$ factor. We see in Fig. \ref{fig:NRCEcs} that Eq. \ref{eq:s-signature} reproduces the energy variation of the baseline of the cross section obtained by the full quantum scattering calculation.
 
The non-thermalized rate coefficient is $K_{\textrm{NRCE}}(E_{\textrm{coll}})= \sigma_{\textrm{NRCE}}(E_{\textrm{coll}}) \times v_{\textrm{coll}}$, with $v_{\textrm{coll}}$ the relative velocity. While the ion velocity distribution is far from following a Maxwell-Boltzmann distribution \cite{hall2013b}, we nevertheless calculate a rate coefficient convoluted with a Maxwell-Boltzmann distribution of the relative velocities $v_{\textrm{coll}}$ of the particle pairs at temperature $T$ for the purpose of comparison,
\begin{equation}
\begin{aligned}
K_{\textrm{NRCE}}(T)=\dfrac{2}{\sqrt{\pi}(k_BT)^{3/2}}\int_0^\infty K_{\textrm{NRCE}}(E_{\textrm{coll}})\sqrt{E_{\textrm{coll}}}e^{-E_{\textrm{coll}}/k_BT}dE_{\textrm{coll}}.
\end{aligned}
\label{eq:knrce}
\end{equation}
The non-thermalized rate is found approximately constant (around $5 \times 10^{-11}$~cm$^{3}$s$^{-1}$) due to the previously observed $E_{\textrm{coll}}^{-1/2}$ slope of the cross section (Fig. \ref{fig:NRCEcs}b). All shape resonances are smoothed out after the thermal average as they are much narrower than the velocity distribution. Our results are consistent within a factor of 2 with the theoretical values of \cite{belyaev2012}. The differences between our model and \cite{belyaev2012} are commented further in the Appendix. At this stage we can say that despite their differences, both models provide results consistent within a factor of $\approx 2$ with the experimental results at 2~K reported in \cite{hall2013b}, or with the measured channel-averaged rate constants in Fig.6a of \cite{hall2013b} assuming a down-scaling by a factor $(1/6)\times (3/4)$ to account for the contribution of the charge exchange resulting from the lowest collision channel (see more details in the Appendix).
 
\begin{figure}
\centering
\includegraphics[width=0.6\textwidth]{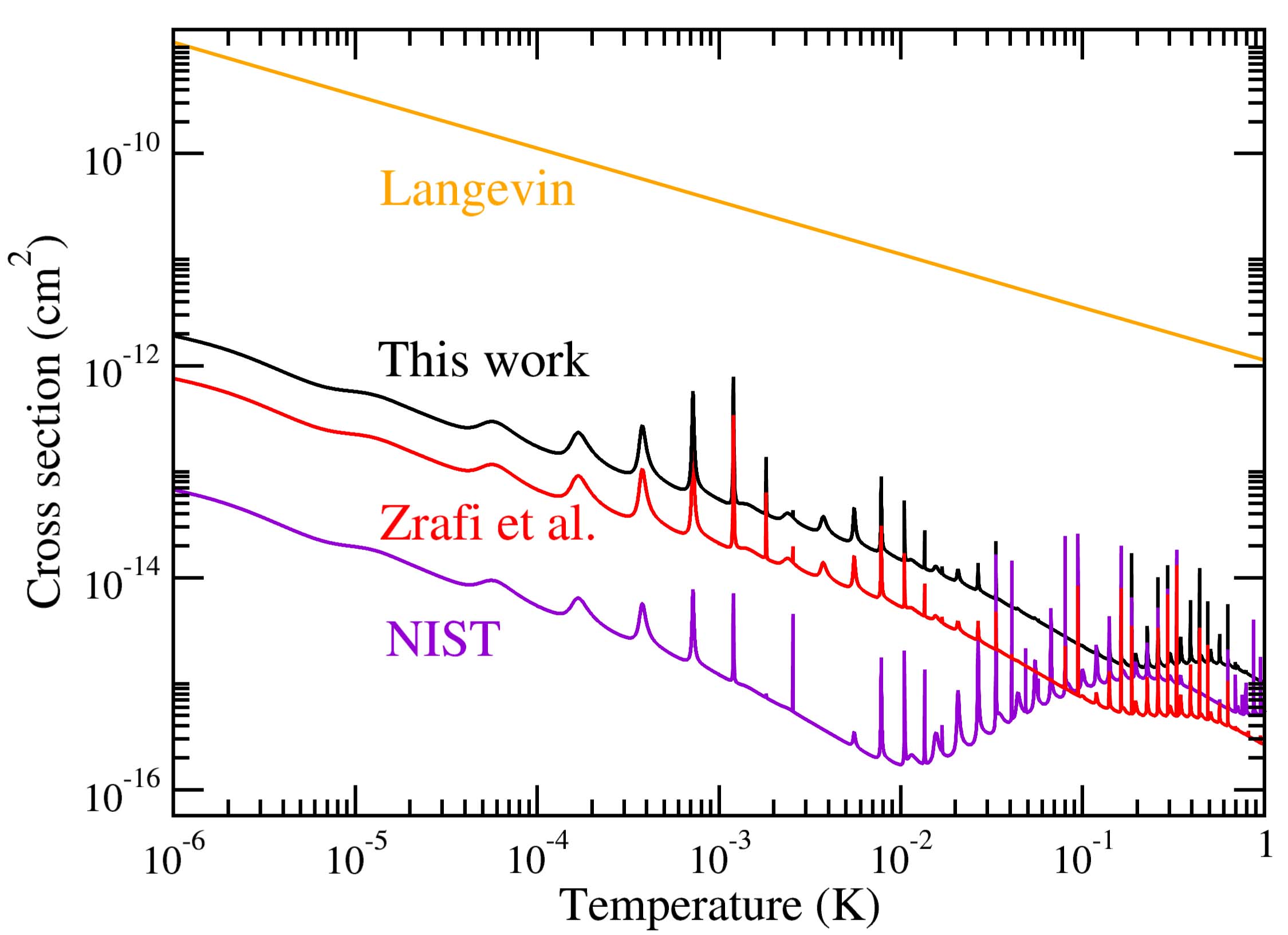}
\caption{NRCE cross sections obtained after the shift of the energy $D_f$ of the dissociation limit of the exit channel $1^3\Pi$ PEC to $D_f$(NIST) (violet), and to $D_f$(\cite{zrafi2020}) (red), compared to the present value (black).  The Langevin cross section $\sigma_L$ (orange) is plotted for comparison. }
\label{fig:NRCE-shift}
\end{figure}

Due to the above ''$s$-wave signature effect'', the issue of the precision of the theoretical model should be taken with great care. For instance, in the present case, we note in the Appendix that the dissociation energy $D_f$ of the exit charge-exchange channel Rb$^+$+Ca($4s4p\,^3P$) is not properly determined compared to the sum of the atomic values extracted from the NIST database \cite{NIST_ASD}. We thus performed two additional calculations of $\sigma_{NRCE}(E_{\textrm{coll}})$, keeping the same entrance channel and SOC, and vertically shifting the $1^3\Pi$ PEC downward to $D_f$(NIST), or upward to $D_f$(\cite{zrafi2020}). As expected, Fig. \ref{fig:NRCE-shift} shows differences larger than one order of magnitude in the $s$-wave regime due to the strong variation of the phase difference between the channels, while the cross section spreads over a factor of $\approx 5$ in the Kelvin range.

\section{Conclusion}

We theoretically investigated the collisional dynamics of laser-cooled Rb atoms and Ca$^+$ ions in the context of the hybrid trap experiment of Ref. \cite{hall2011,hall2012}. We demonstrated that cold RbCa$^+$ ground-state molecular ions are created by radiative association, and that they are protected against photodissociation which would be by black-body radiation and by the Ca$^+$ cooling laser at 397 nm. This study yields a consistent interpretation of the direct observation of RbCa$^+$ ions in the experiment \cite{hall2011,hall2013b}, in contrast with  other hybrid trap experiments using other species (see for instance \cite{mohammadi2021}). Based on novel molecular data, we also confirm that the non-radiative charge exchange, induced by spin-orbit interaction, is a dominant loss process and obtain rates in agreement with experimental observations and a previous calculation \cite{belyaev2012,hall2013b}. Our work also emphasizes on the extreme sensitivity of the model to the accuracy of the molecular data (PECs, spin-orbit couplings). An analogous full quantum scattering treatment including spin-orbit couplings performed in the lab frame is currently in progress for RbSr$^+$ and LiBa$^+$ following the recent experimental results of Ref. \cite{Pascal-thesis}.

\section{Appendix}

\paragraph{Molecular data}

\begin{figure}
\includegraphics[width=11 cm]{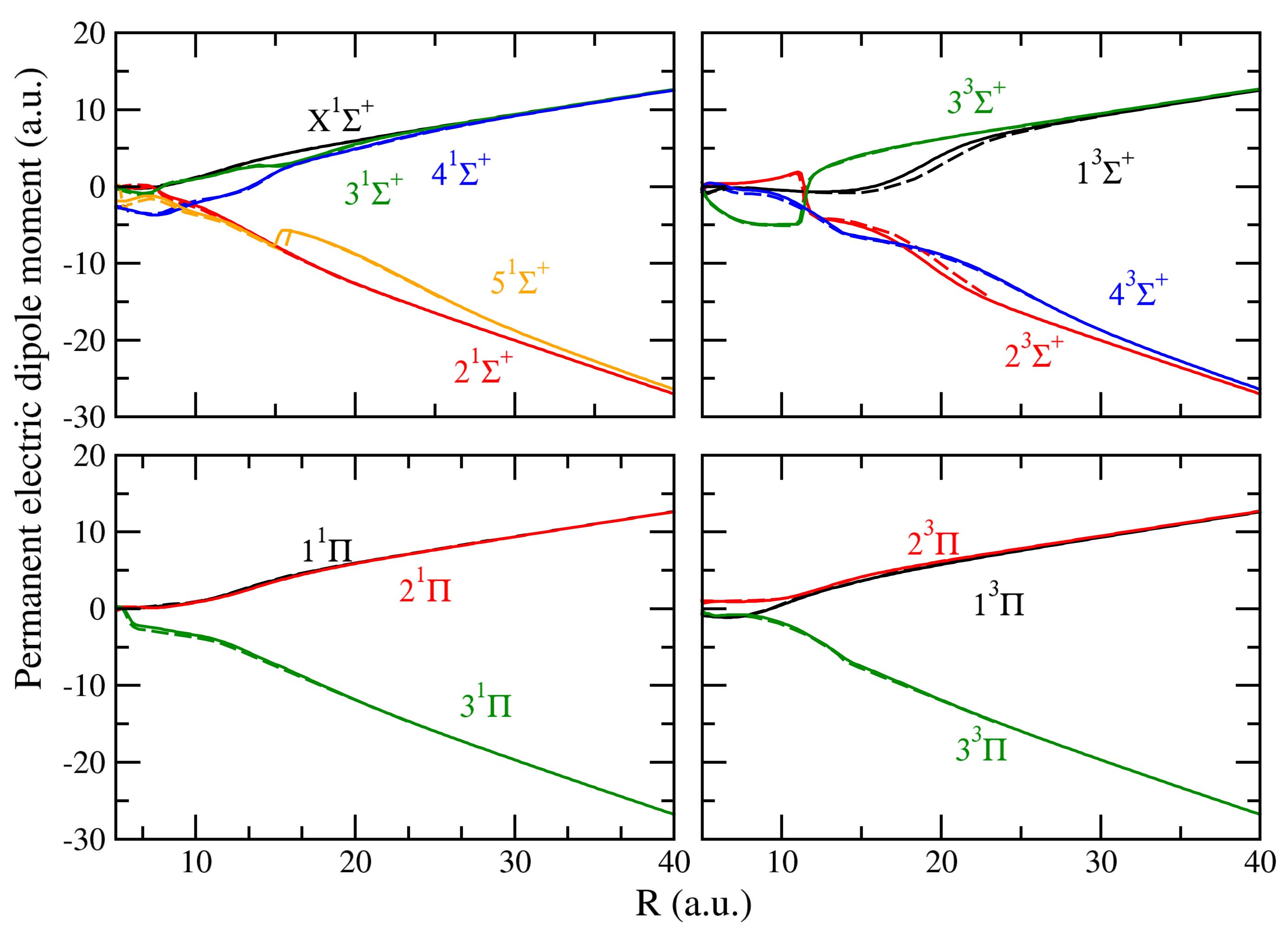}
\caption{Selected PEDMs computed in this work (solid lines) compared to \cite{zrafi2020} (dashed lines).}
\label{fig:PDM} 
\end{figure}

\begin{figure}
\includegraphics[width=9 cm]{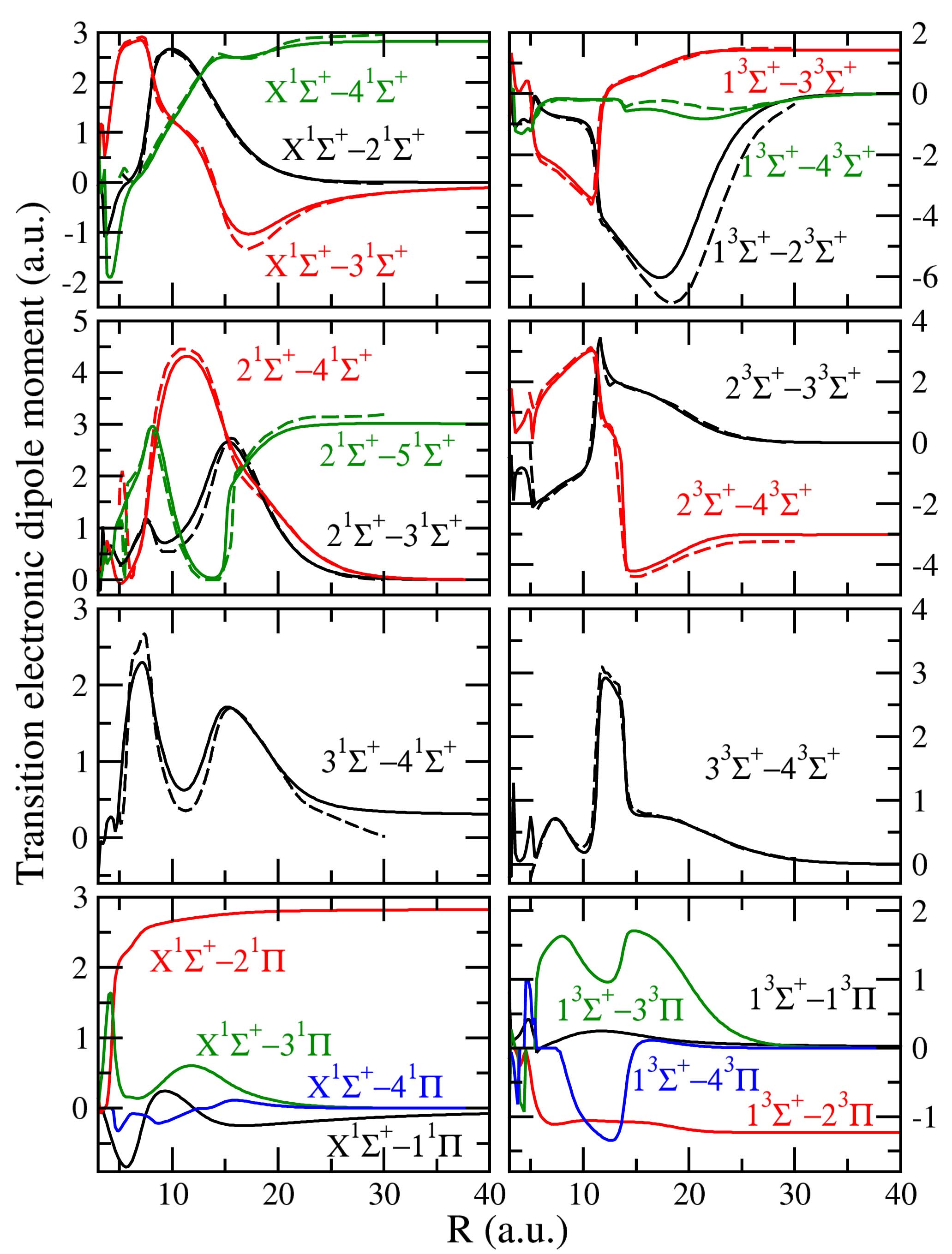}
\caption{ Selected TEDMs computed in this work (solid lines) compared to \cite{zrafi2020} (dashed lines).}
\label{fig:TDM}
\end{figure}

The parameters used are updated from our previous publications \cite{aymar2012,guerout2010}. The optimized basis set for Ca$^+$ is composed of a large set of Gaussian orbitals ($9s$,$8p$,$8d$,$2f$). The cut-off parameters of the core-polarization potential are, in atomic units, ($\rho_s^{Ca^+}=1.89095$, $\rho_p^{Ca^+}=1.6528$, $\rho_d^{Ca^+}=1.827734$), and the ionic core polarizability is $\alpha^{Ca^{2+}}=3.522a_0^3$ \cite{coker1976}. The basis set of Rb is extended with an additional Gaussian $f$-orbital compared to Ref. \cite{guerout2010}. Under such conditions we calculated the energies for the dissociation limits of the RbCa$^+$ PECs which are compared to other values in Table \ref{tab:disslimits}. We see that our basis choice significantly improves the quality of the energy of several atomic states in comparison to \cite{zrafi2020}. Numerous PECs have been calculated, for which we report selected spectroscopic constants in Table \ref{tab:constants} which are compared to the results from \cite{zrafi2020}: Results for the equilibrium distance are in very good agreement, while we found values for the harmonic constant $\omega_e$ larger by about 0.5-1~cm$^{-1}$ than those in \cite{zrafi2020}. The differences in the excitation energy $T_e$ reflect those reported in the dissociation limits. The well depths $D_e$ are in agreement within $\approx 100$~cm$^{-1}$ between the two calculations, corresponding to the typical precision of such calculations. Nevertheless, Figures \ref{fig:PDM} and \ref{fig:TDM} display our results for permanent and transition electric dipole moments, which are quantities sensitive to the structure of the electronic wave functions: we see that the agreement with \cite{zrafi2020} is very satisfactory, confirming that both calculations are very similar.

\begin{table}[ht]
\renewcommand{\arraystretch}{1.6} \addtolength{\tabcolsep}{3.8 pt}
\begin{center}
\begin{tabular}{cccccc}
\hline \hline
Asymptote& \multicolumn{2}{c}{Theory (cm$^{-1}$)} & NIST (cm$^{-1}$)  &\multicolumn{2}{c}{$\delta E$ (cm$^{-1}$)}\\
\cline{2-3} \cline{5-6}
 & This work & \cite{zrafi2020}&\cite{NIST_ASD}  & This work & \cite{zrafi2020}\\
\hline
Rb$^+$($^1$S)+Ca(4s$^21$S)  & -145026& -144904& -145058 & -32& -154\\
Rb$^+$($^1$S)+Ca(4s4p$^3$P)  & -129950& -129624 & -129795 & 155& -171\\
Rb(5s$^2$S)+Ca$^+$(4s$^2$S) & -129444& -129445& -129443 & 1& 2\\
Rb$^+$($^1$S)+Ca(4s3d$^3$D)& -124644& -124136  & -124702 & -58& -566\\
Rb$^+$($^1$S)+Ca(4s3d$^1$D)& -123274& -122611 & -123208 &  66& -597 \\  
Rb$^+$($^1$S)+Ca(4s4p$^1$P)& -121878& -121574& -121405 & 473& 169 \\  
Rb(5p $^2$P)+Ca$^+$(4s$^2$S)& -116707& -116724 & -116705 & 2& 19 \\
Rb(5s$^2$S)+Ca$^+$(3d$^2$D)& -115757 &  & -115756 & 1& \\                         
\hline \hline
\end{tabular}
\caption{Energy of the dissociation limits of RbCa$^+$ PECs with respect to an origin taken at Rb$^+$+Ca$^{2+}$. The energy differences $\delta E$ are calculated as  $\delta E=E(\textrm{NIST})-E(\textrm{this work})$ and $\delta E=E(\textrm{NIST})-E$(\cite{zrafi2020}). }
\end{center}
\label{tab:disslimits}
\end{table}

\newpage

\begin{longtable}[ht]{ccccccccc}
\hline \hline
Asymptote & State & $D_e$   &$R_e$& $B_e$   &$\omega_e$&$\omega_e\chi_e$& $T_e$   & Ref.\\
          &       &cm$^{-1}$&a.u. &cm$^{-1}$&cm$^{-1}$ &cm$^{-1}$       &cm$^{-1}$&
\\ \hline
Rb$^+$($^1$S)+Ca(4s$^2\,^1$S) & 1$^1\Sigma^+$& 3811 & 7.98& 0.034540 & 74.38 & 0.31 & 0 & This work\\
& & 3851 & 7.96 & 0.034600 & 73.02 & -&0 & \cite{dasilva2015}\\
& & 3717 & 8.26 & 0.032371 & 73.43 & 0.50 &0 & \cite{hall2013a}\\
& & 3714 & 7.97 & 0.034735 & 73.53 & 0.35 & 0 & \cite{zrafi2020}\\
& & 3666 & 8.06 & 0.034100 & 73.90 & - & 0& \cite{belyaev2012} \\
& & 3730 & 8.00 & - & - & -& 0& \cite{smialkowski2020}\\
Rb$^+$($^1$S)+Ca($4s4p\,^3$P)   & 1$^3\Sigma^+$ & 7559 & 8.88 & 0.027877 & 65.16& 0.13&11288 & This work\\
& & 7455 & 9.15 & 0.026367 & 77.92 & 0.20 & 10806 & \cite{hall2013a}\\
& & 7494 & 8.90 & 0.027855 & 64.21 & 0.16 & 11503& \cite{zrafi2020}\\
& 1$^3\Pi$ & 3195 & 7.40 & 0.039834& 76.35 & 0.38& 15692 & This work\\
& & 3308 & 7.63 & - & - & - & 18204$^*$ &\cite{zrafi2020} (\cite{hall2013a})\\
& & 3022 & 7.38 & 0.040511 & 75.86 & 0.40 & 15982& \cite{zrafi2020}\\
Rb($5s\,^2$S)+Ca$^+$($4s\,^2$S) & 2$^1\Sigma^+$ & 1249 & 12.88 & 0.013254& 30.24 & 0.15& 18144 & This work\\
& & 1284 & 12.82 & 0.013388 & 28.20 & 0.09 & 17764 & \cite{hall2013a}\\
& & 1272 & 12.86 & 0.010376 & 29.89 & - & 18171& \cite{dasilva2015}\\
& & 1160 & 12.99 & 0.010376 & 29.58 & 0.15& 18015& \cite{zrafi2020}\\
& & 1170 & 13.00 & - & - & - & - & \cite{smialkowski2020}\\
& 2$^3\Sigma^+$ (1$^{st}$ min) & -1727 & 8.32 & 0.031767& 62.14 & 0.23 & 21120 & This work\\
& & -1496 & 8.31 & 0.031976 & - & - & 20128 & \cite{hall2013a}\\
& & -1839 & 8.27 & 0.032261 & 61.54 & 0.49 & 21014 & \cite{zrafi2020}\\
& 2$^3\Sigma^+$ (2$^{nd}$ min) & 181 & 20.02 & 0.005488& 9.85& 0.16& 19212 & This work\\
& & 121 & 17.82 & - & - & - & 19018& \cite{hall2013a}\\
& & 137 & 20.82 & 0.005090 & 8.37 & 0.13&19025& \cite{zrafi2020}\\
Rb$^+$($^1$S)+Ca($4s3d\,^3$D)& 3$^3\Sigma^+$  & 1353 & 11.54 & 0.016521&  106.02 & 5.23& 22840 & This work\\
& & 1481 & 11.42 & 0.016918 & 61.54 & 0.49& 22999&\cite{zrafi2020}\\
& 2$^3\Pi$ & 2894 & 8.69 & 0.029145& 58.41 & 0.21& 21299& This work\\
& & 3027 & 8.67 & 0.029353 & 58.30 & 0.21&21440 &\cite{zrafi2020}\\
& 1$^3\Delta$ & 3226 & 8.38 & 0.031347& 62.58 & 0.22& 20967 & This work\\
& & 3293 & 8.35 & 0.031646 & 61.17 & 0.24&21189 &\cite{zrafi2020}\\
Rb$^+$($^1$S)+Ca($4s3d\,^1$D)& 3$^1\Sigma^+$ & 2943 & 8.71 & 0.028982& 60.68 & 0.22& 22620 & This work\\
& & 3175 & 8.65 & 0.029489 & 62.04 & 0.34& 22835&\cite{zrafi2020}\\  
&  1$^1\Pi$ & 3499 & 8.42 & 0.031054& 58.39 & 0.21& 22064 & This work\\
& & 3542 & 8.43 & 0.031048 & 57.48 & 0.16&22467 &\cite{zrafi2020}\\
& 1$^1\Delta$ & 3981 & 8.32 & 0.031772& 64.62 & 0.19& 21582& This work\\
& & 4184 & 8.29 & 0.032105 & 63.51 & 0.18& 21827&\cite{zrafi2020}\\  
Rb$^+$($^1$S)+Ca($4s4p\,^1$P)& 4$^1\Sigma^+$ & 973 & 15.01 & 0.009756& 19.42 & 0.19& 25986& This work \\
& & 1016 & 14.89 & 0.009952 & 22.61 & 0.31& 26035 &\cite{zrafi2020}\\  
& 2$^1\Pi$  & 1923 & 8.46 & 0.030752& 51.10 & 0.38& 25036 & This work\\
& & 1575 & 8.47 & 0.030755 & 46.14 & 0.51& 25476&\cite{zrafi2020}\\
Rb($5p\,^2$P)+Ca$^+$($4s\,^2$S)& 5$^1\Sigma^+$ (1st min)  & -350 & 9.59 & 0.023907& 50.28 & 1.53 &32480 & This work\\
& 5$^1\Sigma^+$ (2ed min)& 1216 & 20.26 & 0.005343& 20.29 & 0.05 & 30914& This work\\
& & 2051 & 13.61 & 0.011912& 16.90 & 0.31 & 30754& \cite{zrafi2020}\\
& 4$^3\Sigma^+$  & 1801 & 18.50 & 0.006404& 18.53 & 0.04& 30329& This work\\
& & 1745 & 18.57 & 0.006398 & 20.67 & 0.04&31163 &\cite{zrafi2020}\\
& 3$^1\Pi$  & 275 & 12.73 & 0.013571& 16.24 & 0.52 & 31855 & This work\\
& & 131 & 17.00 & 0.007635 & 8.70 & 0.07& 31782&\cite{zrafi2020}\\
& 3$^3\Pi$ (1st min) & 233 & 13.48 & 0.012109& - & - & 31897& This work\\
& 3$^3\Pi$ (2ed min) & 281 & 15.60  & 0.009032& 14.82 & 0.34& 31849& This work\\
& & 202 & 16.19 & 0.008418 & 12.17 & 0.16 &31718 &\cite{zrafi2020}\\  
Rb($5s\,^2$S)+Ca$^+$($3d\,^2$D)& 6$^1\Sigma^+$   & 685 & 15.40 &  0.009268 & 55.72 & 3.13&32395 & This work \\
& 5$^3\Sigma^+$  & 937 & 14.15 & 0.010978& 53.83 & 2.07&32143 & This work\\
& 4$^1\Pi$  & 678 & 14.96 & 0.009827& 25.87 & 0.33& 32402& This work\\
& 4$^3\Pi$  & 997 & 14.00 & 0.011228 & 39.73 & 1.05& 32083 & This work\\
& 2$^1\Delta$  & 1015 & 13.59 & 0.011901  & 27.39 & 0.16&32065 & This work\\
& 2$^3\Delta$  & 1162 & 13.16 & 0.012701 & 29.36 & 0.16& 31918& This work \\       
\hline \hline
\caption{Computed spectroscopic constants of RbCa$^+$. Equilibrium interatomic distances $R_e$, well depths $D_e$, transition energies $T_e$, harmonic constants $\omega_e$, anharmonic vibrational frequencies $\omega_e\chi_e$, and rotational constants $B_e$. Note that the $T_e$ value for the 1$^3\Pi$ PEC marked with a $*$ symbol is actually copied from \cite{zrafi2020}, where it is most probably improperly displayed.}
\label{tab:constants}
\end{longtable}

\paragraph{Spin-orbit coupling matrices}
To describe SOC in the Rb$^+$ + Ca($4s4p\,^3P$) limit, the good Hund's case c quantum number $|\Omega|$ could be $0^+,0^-,1$. Eventually, the total potential energy matrices of Rb($5s\,^2S$) $+$ Ca$^+$($4s\,^2S$) and Rb$^+$ + Ca($4s4p\,^3P$) including SOC are expressed as
\begin{equation}
 H(\Omega=0^- ) = \left(\begin{array}{ccc}
    V(2^3\Sigma^+)& A_{12} & 0 \\
    A_{21}& V(1^3\Pi)-A_{22} & A_{23} \\
    0 & A_{32} & V(1^3\Sigma^+) 	
\end{array}\right),
\end{equation}
\begin{equation}
 H(\Omega=0^+ ) = \left(\begin{array}{ccc}
     V(2^1\Sigma^+) & A'_{12} \\
     A'_{21} & V(1^3\Pi)-A'_{22} 	
\end{array}\right),
\end{equation}
\begin{equation}
 H(|\Omega|=1) = \left(\begin{array}{ccc}
     V(2^3\Sigma^+)& A''_{12} & 0 \\
     A''_{21}& V(1^3\Pi) & A''_{23} \\
     0 & A''_{32} & V(1^3\Sigma^+) \\
\end{array}\right),
\label{eq:soc1}
\end{equation}
where the $R$-dependent coupling terms $A_{ij}, A'_{ij}, A''_{ij},$ are displayed in Fig. \ref{fig:SOC}. Note that we neglected in the $|\Omega|=1$ matrix (Eq.\ref{eq:soc1}) the contribution of the Rb$^+$ + Ca($4s4p\,^1P$) limit, which is located about 8000~cm$^{-1}$ above the Rb$^+$ + Ca($4s4p\,^3P$) one. The consequence is that after diagonalization, the energy of  Rb$^+$ + Ca($4s4p\,^1P$) is slightly shifted compared to Rb$^+$+ Ca($4s4p\,^3P_{0,2}$) by 0.518~cm$^{-1}$, which can be neglected in the rest of the study.

\begin{figure}
\centering
\includegraphics[width=14 cm]{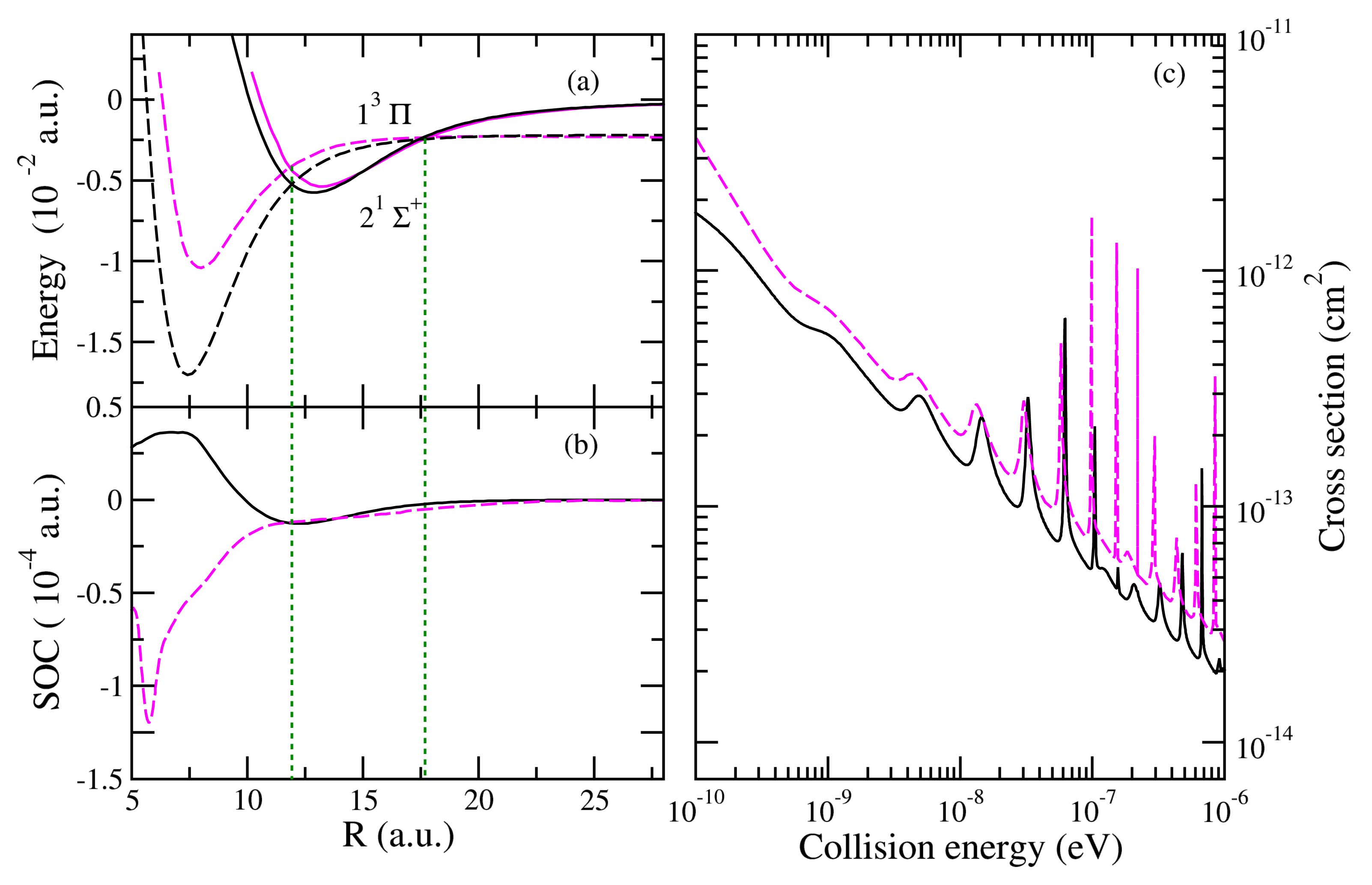}
\caption{This work (black) and Ref. \cite{belyaev2012} (magenta): (a) Hund' case a PECs of the $1^3\Pi$ (solid line) and $2^1\Sigma^+$ (dashed line) states,  the vertical dashed lines indicate the two crossings; (b) SOC between the $1^3\Pi$ and $2^1\Sigma^+$ states, the vertical dashed lines mark the values of SOC at the two crossings; (c) The calculated NRCE cross sections as a function of the collision energy.}
\label{fig:NRCE-theory}
\end{figure}

\paragraph{Differences between our model and the one of \cite{tacconi2011}}
To understand the differences between theoretical models, a detailed comparison between the results of this work and Ref. \cite{tacconi2011} is presented in Figure \ref{fig:NRCE-theory}. We can see that the dissociation energies of $1^3\Pi$ and $2^1\Sigma^+$ are close, but the potential wells differ. Although the SOCs trends are quite different over the short-range $<$ 10 au, they are similar both at the intersections and at long-range. The patterns of the cross section are also consistent as they display a comparable density of shape resonances. Although the difference in the SOC is large at short range, this good agreement also demonstrates that the short-range SOC does not play a role here.

\paragraph{Comparison with experimental data}
We reported in Fig.\ref{fig:NRCEcs}b  ''down-scaled experimental data'' that we obtained by multiplying the measured rate constants in Fig.6a of \cite{hall2013b} by a factor $(1/6)\times (3/4)$. This scaling is reasoned from the following two assumptions reported in \cite{hall2013b}: 
\begin{itemize}
    \item the reported experimental data corresponded to the total reaction rate to which the contributions of the Ca$^+$($4s$) and Ca$^+$(4p) channels were estimated to be $\approx$ 1:5. 
    \item the ratio of the molecule-formation vs. charge-transfer rates was roughly estimated to be $\approx$ 1:3.
\end{itemize}

Therefore, a scaling factor of $1/6 \times 3/4 = 1/8$ yielded a closer match to the theoretical results shown in Fig~\ref{fig:NRCEcs}(b). It needs to be emphasised that the channel and product branching ratios reported in \cite{hall2013b} are rough estimations, and consequently also the relevant scaling factors adapted here need to be treated with caution. Still, an overall scaling of $1/8$ of the channel averaged rate constants in \cite{hall2013b} seems to yield an acceptable agreement with the presented theoretical values  in the lowest collision channel reported in Fig.~\ref{fig:NRCEcs}b.

\section{Acknowledgments}
X. X. acknowledges support from the Chinese Scholarship Council (Grant No. 201706240178), and support from COST Action CA17113 ''Trapped Ions: Progress in Classical and Quantum Applications'' for funding a Short Term Scientific Mission enabling the present work. X.X. is thankful to Dr. Alexander D. D\"orfler for enlightening discussions in Basel. S.W. acknowledges financial support from the Swiss National Science Foundation, grant nr. $00020\_175533$.

%

	
\end{document}